\documentclass[12pt]{iopart}
\usepackage{iopams} 
\usepackage{bm,color,bbm}
\usepackage{ulem} 
\expandafter\let\csname equation*\endcsname\relax
\expandafter\let\csname endequation*\endcsname\relax   
\usepackage{euscript}
\usepackage{hyperref, mathtools,graphicx,amsmath,cases}
\usepackage{amssymb,amsthm}
\newcommand{\beq}{\begin{equation}}
\newcommand{\eeq}{\end{equation}}
\newcommand{\nn}{\nonumber}
\newcommand{\id}{\mathbbm{1}}
\newtheorem{theorem}{Theorem}

\newtheorem{propos}{Proposition}
\newtheorem{corollary}{Corollary}
\newtheorem{lemma}{Lemma}

\definecolor{bgr}{rgb}{0.0, 0.26, 0.15}
\newcommand{\B}{{\mathcal B}}
\newcommand{\str}{\mathcal{R}}
\def\half{\frac{1}{2}}
\newcommand{\orcid}[1]{\href{https://orcid.org/#1}{\resizebox{10px}{!}{\includegraphics{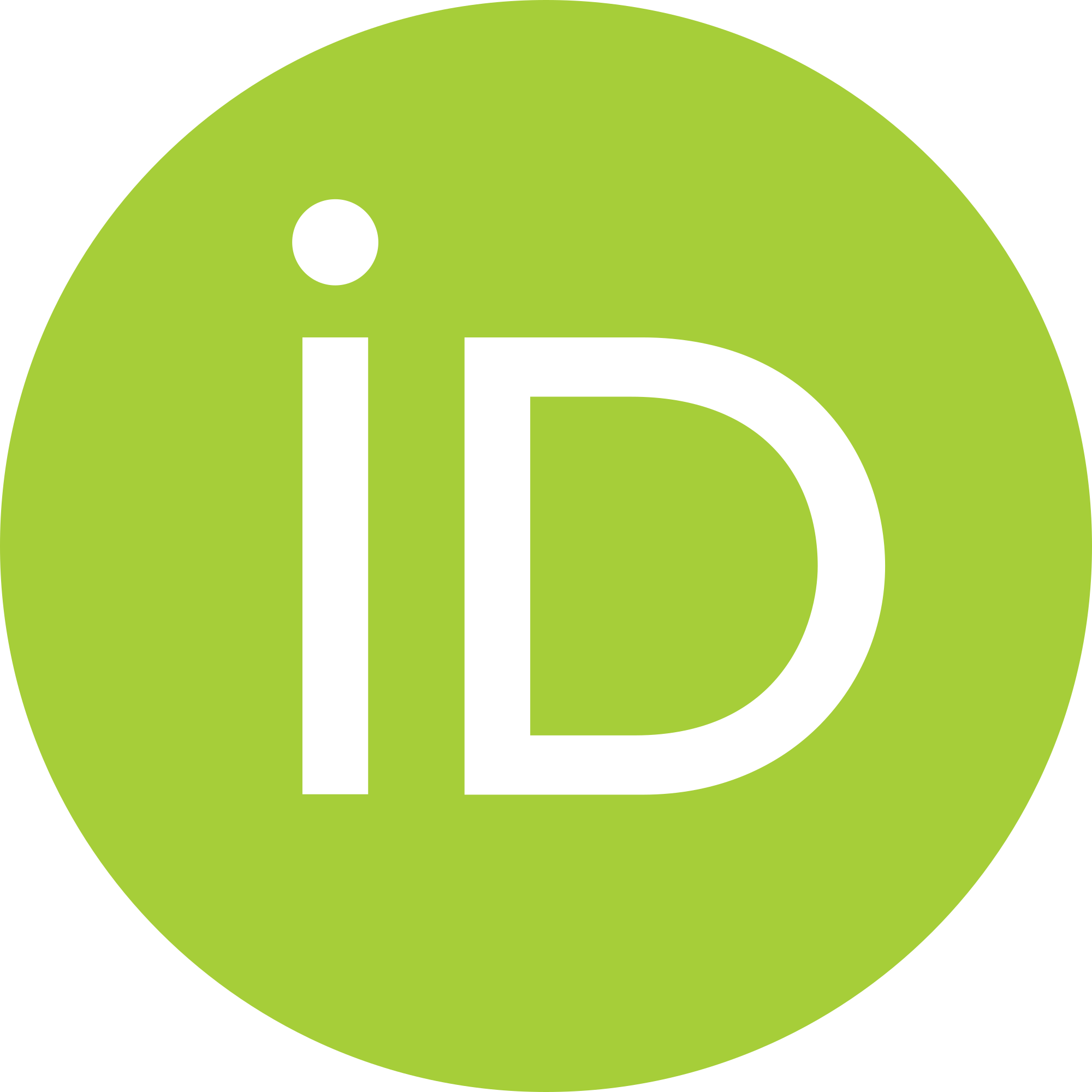}}}}

\begin{document}

\title{Mermin and Svetlichny inequalities for non-projective measurement observables}

\author{Mohd Asad Siddiqui$^1$\orcid{0000-0001-5003-7571} and Sk Sazim$^{2}$\orcid{0000-0003-3117-0785}}
\address{$^1$  Institute of Fundamental and Frontier Sciences, University of Electronic Science and Technology of China, Chengdu 610051, People's Republic of China}
\address{$^2$ RCQI, Institute of Physics, Slovak Academy of Sciences, 845 11 Bratislava, Slovakia}
\ead{\href{mailto:asad@ctp-jamia.res.in}{asad@ctp-jamia.res.in} and \href{mailto:sk.sazimsq49@gmail.com}{sk.sazimsq49@gmail.com}}
\vspace{10pt}

\begin{abstract} 
The necessary and sufficient criteria for violating the Mermin and Svetlichny inequalities by arbitrary three-qubit states are presented.  
Several attempts have been made, earlier, to find such criteria, however, those extant criteria are neither tight for most of the instances, nor fully general.   
We generalize the existing criteria for Mermin and Svetlichny inequalities which are valid for the local projective measurement observables as well as for the arbitrary ones.
We obtain the maximal achievable bounds of the Mermin and Svetlichny operators with unbiased measurement observables for arbitrary three-qubit states and with arbitrary observables for three-qubit states having maximally mixed marginals. We find that for certain ranges of measurement strengths, it is possible to violate Mermin and Svetlichny inequalities only by biased measurement observables. The necessary and sufficient criteria of violating any one of the six possible Mermin and Svetlichny inequalities are also derived. 
\end{abstract}

%
\vspace{2pc}

\vspace{4pc}

\maketitle
%
%

\section{Introduction}

Bell inequalities play a pivotal role in demarcating the correlations entertained by two or more distant quantum particles than admissible by their classical counterpart \cite{PhysicsPhysiqueFizika.1.195}. The most celebrated Bell inequality is the Bell-CHSH inequality for two-qubit states \cite{PhysRevLett.23.880}.  Its straightforward generalization to three qubits is popularly known as Mermin \cite{PhysRevLett.65.1838} and Svetlichny \cite{PhysRevD.35.3066} inequalities. At their inception, these inequalities are considered for sharp dichotomic measurement observables, i.e., observables with two sharp effects corresponding to two distinct outcomes, say $\pm 1$ (known as projective measurement observables).

However, sometimes, an observer might not be able to measure projective observables, due to the apparatus limitations such as detector noise or interactions with the environment. Also, non-projective measurements are not only theoretically intriguing but has many potential applications in quantum information processing protocols, e.g., most importantly, non-orthogonal state discrimination and its crucial role in randomness extraction plus quantum cryptography \cite{heinosaari_ziman_2011}.
To deal with such situations, Hall and Cheng have provided the necessary and sufficient conditions to violate the Bell-CHSH inequalities for non-projective measurements on arbitrary two qubits \cite{Hall_2022}, by improving the Horodecki bound (of sharp observables) \cite{HORODECKI1995340}.
The generalization of the bound is also useful for the task of `resource recycling' where bonafide parties use the noisy detector to implement the task \cite{PhysRevA.104.L060201,PhysRevA.105.022411}. Here, by resource, we mean any quantum correlations which can be shared by two or more parties, e.g., quantum entanglement, nonlocality, steering, etc \cite{RevModPhys.81.865,RevModPhys.86.419}. The framework provided by Hall and Cheng in Ref. \cite{Hall_2022}, has inspired us to do the similar analysis for three qubit Bell inequalities.

Finding analytical solutions for the maximum value of the bipartite as well as the multipartite Bell operators for arbitrary quantum states is of great importance, specifically, it helps to capture the deviation of quantum correlations from the classical ones \cite{HORODECKI1995340,2001JPhA...34.6043S,2020JPhA...53d5303L}. This will also entail a necessary and sufficient criterion for violating the respective Bell inequalities \cite{HORODECKI1995340}. Moreover, the degree of violating Bell inequalities in a given setup is a key ingredient for many tasks, such as bounding key rates of secure cryptographic protocols \cite{2006NJPh....8..126A} and the certification of random number generators \cite{2010Natur.464.1021P}, to name a few. Therefore, the search of finding such an upper bound is still going on. The quantum upper bound for Mermin and Svetlichny inequalities were proposed earlier, respectively in Refs. \cite{Siddiqui2019, PRXQuantum.2.010308} and Ref. \cite{PhysRevA.96.042323}, however, in both the cases the observables under consideration were sharp. Moreover, those proposed bounds were not always tight for most of the situations.

In this work, we consider the Mermin and Svetlichny operators for non-projective quantum observables and find their maximal value for arbitrary three qubit states. Our analysis gives more insights into the tightness of these bounds by generalizing the previous bounds. 
 In particular, we find a new generalization of the upper bound of Svetlichny inequality for projective measurement which is tight for wider classes of three qubit states. Our results are particularly useful in the cases where there are apparatus limitations such as detector noise. Also, these bounds are useful for the task where the preservation of entanglement is paramount, such as recycling resources \cite{PhysRevA.104.L060201,PhysRevA.105.022411,PhysRevLett.114.250401,math4030048,PhysRevLett.125.090401}, randomness generations \cite{PhysRevA.95.020102} and state discriminations \cite{PhysRevLett.126.180502}.

In the following section, we describe the three qubit Bell inequalities in a nutshell for dichotomic quantum observables. Also, we introduce the framework of general measurement (i.e. non-projective) observables with a concept like `bias' and `strengths' \cite{Hall_2022}. In Sec. \ref{Mer_unbias}, we find the necessary and sufficient condition(s) of violating Mermin inequalities by finding the upper bound of Mermin operators for general measurement observables for arbitrary three qubit states. We did a similar analysis for Svetlichny operators in Sec. \ref{Svet_unbias}. Finally, we conclude in Sec. \ref{concl}.

\section{Bell inequalities for three qubits} \label{main}
The general three qubit state $\rho_{ABC}$ in   
$\mathcal{L}(\mathcal{H}_A,\mathcal{H}_B ,\mathcal{H}_C) $ can be expressed as
\beq 
\rho=\frac{1}{8}\sum_{\mu,\nu,\gamma=0}^3\Lambda_{\mu\nu\gamma}
\sigma_{\mu}\otimes\sigma_{\nu}\otimes \sigma_{\gamma}, \label{3qstate}
\eeq
where $\Lambda_{\mu\nu\gamma}={\rm Tr}[(\sigma_{\mu}\otimes\sigma_{\nu}\otimes \sigma_{\gamma})\rho]$. 
The coefficient $\Lambda_{000}=1$ is the normalization condition; 
$\bm{l}=\{\Lambda_{i00};i=1,2,3\}$, $\bm{m}=\{\Lambda_{0j0};j=1,2,3\}$, $\bm{n}=\{\Lambda_{00k};k=1,2,3\}$ are 
the bloch vectors for three parties respectively; $ \Theta=[\Lambda_{ij0}]$, $ \Phi=[\Lambda_{i0k}]$, 
$ \Omega=[\Lambda_{0jk}]$ are the two party correlation matrices 
and $T=[\Lambda_{ijk}]$ is the tripartite correlation matrix. 

Any general qubit observable $X$ with two outcomes $\omega=\pm 1$ can be described by two effects $\{X_+, X_-\}$ with $X_{\pm}\geq 0$ and $X_++X_-=\mathbbm{1}$. For projective effects, $X_{\pm}^2=X_{\pm}$, and without loss of generality, one can show that $X=\bm x\cdot\bm \sigma$, where $\bm{x} =\{x_1, x_2,x_3\}$ is three-dimensional real unit vector and $\bm{\sigma}=(\sigma_{1},\sigma_{2},\sigma_{3})$ as a vector of Pauli spin matrices. Now, the tripartite correlation can be established for projective observables $X,Y,Z$ as  $XYZ = (\bm{x}\cdot\bm{\sigma})\otimes(\bm{y}\cdot\bm{\sigma})\otimes(\bm{z}\cdot\bm{\sigma})$.
By taking linear combinations of these correlations one can define the following identities for observables $X,X',Y,Y',Z,Z'$
\begin{align} \label{merm}
\mathcal{E}=XYZ' +  XY'Z+ X'YZ -  X'Y'Z',\\ 
\mathcal{E'}= X'Y'Z + X'YZ'+ XY'Z' -  XYZ.\nn
\end{align}
The expectation value of operator, $\mathcal{E}$ for state $\rho$ is given by $\langle\mathcal{E}\rangle={\rm Tr}[\mathcal{E} \rho]$. We define the expectation value of Mermin and Svetlichny operators respectively as
$\EuScript{M}=|\langle{\mathcal E}\rangle|$ and
$\EuScript{S}=|\langle{\mathcal E-\mathcal E}'\rangle|$,
where the values depend both on the state and observable parameters. 
The inequalities,
\begin{align}\label{form-svet-marm}
\EuScript{M}:=|\langle{\mathcal E}\rangle|\leq 2, \qquad {\rm and}\qquad
\EuScript{S}:=|\langle{\mathcal E-\mathcal E}'\rangle|\leq 4,
\end{align} 
are known in literature as Mermin inequality \cite{PhysRevLett.65.1838}, 
and Svetlichny inequality \cite{PhysRevD.35.3066} respectively. The violation of these inequalities admits tripartite Bell-nonlocality, whereas it's genuine nonlocality for the later one \cite{PhysRevA.81.052334}. The attempts to find optimal values of Mermin and Svetlichny operators have been done earlier in Refs. \cite{Siddiqui2019,PRXQuantum.2.010308,PhysRevA.96.042323}, but only for sharp observables. To recall, we state these results below:
\begin{lemma}\label{theoremold1}
Grasselli et al \cite{PRXQuantum.2.010308} The maximum quantum value of Mermin operator on three qubit states for projective observables  $X$, $X'$, $Y$, $Y'$, $Z$ and $Z'$, is given by  
\beq 
\EuScript{M} \leq M(T):= 2\sqrt{s^2_1(T) + s^2_2(T)},\nn 
\eeq
where $s_1(T), s_2(T)$ are the two largest singular values of correlation matrix $T$ (i.e., square roots of the eigenvalues of $T^\top T$).
\end{lemma}
Note that the upper bound in Lemma \ref{theoremold1} is tight only if the correlation matrix $T$ of the considered state meets certain conditions, i.e., for the specific choices of global vectors, it should satisfy
\begin{align}\label{cond-mermsvet}
\bm y\otimes \bm z'+\bm y'\otimes \bm z=2\sqrt{s^2_1(T)\big/[s^2_1(T)+s^2_2(T)]}\bm s_1,\nn\\ {\rm and}\quad \bm y\otimes \bm z-\bm y'\otimes \bm z'=2\sqrt{s^2_2(T)\big/[s^2_1(T)+s^2_2(T)]}\bm s_2,
\end{align} 
where $([\bm y, \bm y'], [\bm z,\bm z'])$ and $(\bm s_1, \bm s_2)$ denotes the local measurement vectors of two distant parties, and the normalized nine-dimensional eigenvectors of matrix $T^\top T$ respectively \cite{PRXQuantum.2.010308}. The above results of Mermin inequality can further be generalized for non-projective ones, which is one of the objective of this work.

Next, we recall the result for Svetlichny inequality: 
\begin{lemma}\label{theoremold2}
Ming et al \cite{PhysRevA.96.042323} The maximum quantum value of Svetlichny operator on three qubit states for projective observables  $X$, $X'$, $Y$, $Y'$, $Z$ and $Z'$, is given by  
\beq 
\EuScript{S} \leq S(T):= 4s_{\max}(T), \nn
\eeq
where $s_{\max}(T)$ is the largest singular value of correlation matrix $T$. 
\end{lemma}
Here, it is evident that the bound in Lemma \ref{theoremold2} is achievable for three qubit states if two conditions, i.e., i) largest singular value of $T$, $s_{\max}(T)$ has degeneracy two, and ii) the conditions in Eq. (\ref{cond-mermsvet}) for $s_1(T)=s_2(T)=s_{\max}(T)$, are satisfied simultaneously. This is only possible for some $GHZ$-class of states with projective measurement \cite{PhysRevA.96.042323}.
However, these results of Svetlichny inequality with projective measurements can be improved further, and can also be generalized for non-projective ones. 

A general qubit observable $X$ can be be decomposed as \cite{Hall_2022}
\beq 
X =  \B  \mathbbm{1} + \str\bm \sigma\cdot\bm x, \label{observable}
\eeq
where  $\B$ is the {\it bias} of the observable; and $\str\geq0$ is {\it strength} (sharpness) parameter; which satisfy the constraint
 \beq \label{constraint}
 \str + |\B| \leq 1,
 \eeq 
where $|\bm a|:=\sqrt{\bm a\cdot\bm a}$. An unbiased observable with $\B=0$ has maximum sharpness $\str=1$ corresponds to projective one, while with minimum strength $\str=0$
corresponds to trivial observable  $X=\B\id$, equivalent to tossing a coin with outcome probabilities $\half(1\pm\B)$.

The expectation value of $XYZ$ is thus given by
\begin{align}
\langle XYZ \rangle & ={\rm Tr}[(X\otimes Y\otimes Z) \rho] \nn\\
&=\B_X \B_Y \B_{Z}+\B_Y \B_{Z} \str_X \sum_{i=1}^{3}  x_i \Lambda_{i,0,0} +\B_X \B_{Z} \str_Y \sum_{j=1}^{3}  y_j \Lambda_{0,j,0} \\&+ \B_X \B_Y  \str_{Z}  \sum_{k=1}^{3}  z_k \Lambda_{0,0,k} \nn  +\B_{Z} \str_X \str_Y \sum_{i,j=1}^{3}  x_i   y_j\Lambda_{i,j,0}+\B_Y \str_X \str_{Z} \sum_{i,k=1}^{3}  x_i   z_k\Lambda_{i,0,k}\\&+\B_X \str_Y \str_{Z} \sum_{j,k=1}^{3}  y_j   z_k\Lambda_{0,j,k}\nn+\str_X \str_Y \str_{Z}  \sum_{i,j,k=1}^{3}  x_i   y_j z_k \Lambda_{i,j,k}, \nn\\
& =\B_X \B_Y \B_{Z}+\B_Y \B_{Z} \str_X \bm l^{\top}  \bm x+\B_X \B_{Z} \str_Y \bm m^{\top}  \bm y +\B_X \B_Y  \str_{Z} \bm n^{\top} \bm z+\B_{Z} \str_X \str_Y \bm x^{\top} \Theta  \bm y\nn\\&
+\B_Y \str_X \str_{Z} \bm x^{\top}  \Phi \bm z+\B_X \str_Y \str_{Z} \bm y^{\top} \Omega \bm z+ \str_X \str_Y \str_{Z} \bm x^{\top} T(\bm y\otimes \bm z),\nn
\end{align}
for $X=\B_X\id+\str_X\bm\sigma\cdot\bm x$, $Y=\B_Y\id+\str_Y\bm\sigma\cdot\bm y$ and $Z=\B_Z\id+\str_Z\bm\sigma\cdot\bm z$. Using these generalize tripartite correlations, we will evaluate the admissible quantum upper bounds of Mermin and Svetlichny inequalities.

\section{Generalizing the Mermin bound}\label{Mer_unbias}
We evaluate the quantum upper bound of Mermin inequality using general observables defined in Eq. (\ref{observable}). Let us consider the unbiased observables with $\B=0$, then 
the expectation value of Mermin operator (Eq. (\ref{merm})) becomes
\begin{align} \label{mermbzero}
\langle \mathcal{E} \rangle_{ub} =& \str_X \bm x^{\top} T\left[(\str_Y\bm y\otimes \str_{Z'}\bm z')+  (\str_{Y'}\bm y'\otimes \str_Z \bm z)\right]+ \str_{X'} \bm x'^{\top}T\big[(\str_Y\bm y\otimes \str_Z \bm z)\nn \\&- ( \str_{Y'}\bm y'\otimes \str_{Z'}\bm z')\big].
\end{align}
The quantum upper bound of the Mermin operator has been evaluated for unbiased observables and compactly stated in the following theorem.
\begin{theorem}\label{thm3}
The tight upper bound of Mermin inequality on three qubit states  for unbiased observables $X,X',Y,Y',Z,Z'$, with strengths $\str_X, \str_{X'}, \str_Y,\str_{Y'},\str_Z, \str_{Z'}$  and correlation matrix $T$ is given by 
\begin{eqnarray} \label{bound}
 \EuScript{M} &\leq& \EuScript{M}_0:= \sum_{i=1}^{2} s_i(T) s_i(V), \nn \\ 
 &=&\half  [s_1(T)+s_2(T)]I_+(V) +  \half  [s_1(T)- s_2(T)]I_-(V) ,\qquad
\end{eqnarray}
where $V$ is a $3\times 9$ matrix defined in \ref{appen}. And $s_1(T),s_2(T)$  and $s_1(V), s_2(V)$ are the two largest singular values of $T$ and $V$ respectively, and $I_\pm(V):=s_1(V)\pm s_2(V)\geq0$, may be calculated using
\begin{align}\label{iplusminus1}
	I^2_\pm(V)&=I_0+2\biggr[  I_{XY}^Z\cos\theta_x \cos\theta_y + I_{XZ}^Y\cos\theta_x \cos\theta_z+ I_{YZ}^X\cos\theta_y \cos\theta_z\pm\str_X \str_{X'}\sin\theta_x \nn\\&\times\Big
\{I_{YZ}^0\sin^2{\theta_y}+ I_{ZY}^0 \sin^2{\theta_z}+  I_1^2 (1- \cos 2\theta_y \cos 2\theta_z) \Big\}^{\frac{1}{2}}\biggr],
\end{align}
where $I_0=\str_X^2(\str_Y^2\str_{Z'}^2+\str_{Y'}^2\str_Z^2)+\str_{X'}^2(\str_Y^2\str_Z^2+\str_{Y'}^2\str_{Z'}^2)$, $I_{ij}^k=\str_i \str_{i'} \str_j \str_{j'} (\str_k^2-\str_{k'}^2)$, $I_{ij}^0=\str_i^2\str_{i'}^2(\str_j^4+\str_{j'}^4)$, $I_1=\str_Y \str_{Y'} \str_Z \str_{Z'}$, $\cos\theta_x=\bm x\cdot\bm x'$, $\cos\theta_y=\bm y\cdot\bm y'$, $\cos\theta_z=\bm z\cdot\bm z'$ and $\{i,j,k\}\in [X,Y,Z]$.

\end{theorem}
We prove the Theorem \ref{thm3} in \ref{appen}. It can be noticed from Theorem \ref{thm3} that the Mermin inequality can be violated by measuring unbiased observables (with given strengths and relative angles), if and only if $\EuScript{M}_0>2$. Further, Theorem \ref{thm3} generalizes the special case represented in Lemma \ref{theoremold1}, valid for all three qubit states. Note that the upper bound in the theorem is invariant under local unitary transformations on every party as these transformations leave measurement strength, relative angles, and singular values invariant (similar reasoning to Ref. \cite{Hall_2022}). Recently, a tight upper bound of Mermin inequality was found for sharp observables in Ref. \cite{PRXQuantum.2.010308} (see Lemma \ref{theoremold1}). The link between Lemma \ref{theoremold1} and Theorem \ref{thm3} is evident by following simplified analysis.
Using Cauchy–Schwarz inequality and the identity ${\rm Tr}[V^\top V]=\sum_{i=1}^{2} s^2_i(V)$ in Eq. (\ref{bound}), we get
\begin{align*} 
\EuScript{M}^2\leq  {\rm Tr[V^\top V]} \sum_{i=1}^{2} s^2_i(T)\leq  M^2(T).
\end{align*}
The above inequality is achieved, as ${\rm Tr[V^\top V]}\leq \max\{A^2,B^2,C^2,D^2\}\leq 4$, where $A, B, C,$ and $D$ are defined in \ref{appen}. Therefore, the bound in Lemma \ref{theoremold1} is the upper bound of the Mermin operator for unbiased observables.

A more general bound can be obtained from Theorem \ref{thm3} in the following. 
\begin{corollary} \label{corollary1}
The tight upper bound of Mermin inequality on three qubit states for unbiased observables $X,X',Y,Y',Z,Z'$, with strengths $\str_X=\str_{X'}$, $\str_Y=\str_{Y'}$,  $\str_Z$=$\str_{Z'}$   and correlation matrix $T$ is given by	
	\beq \label{corollary1eq}
\EuScript{M}\leq 2\str_X \str_Y \str_Z\sqrt{s^2_1(T)+s^2_2(T)}.
	\eeq
	The above bound is achieved for any relative angle satisfying
	\beq \label{relan}
	\sin \theta_x \sqrt{1-\cos^2 \theta_y \cos^2 \theta_z} = \frac{2s_1(T)s_2(T)}{s^2_1(T)+s^2_2(T)}.
	\eeq
\end{corollary}

The proof of above corollary is given in \ref{appen}. If the largest singular value of $T$ has degeneracy two (at least), and say, it is $s_{\max}(T)$, then Eq. (\ref{corollary1eq}) becomes
	\beq \label{degenerate}
\EuScript{M}  \leq 2\sqrt{2} s_{\max}(T) \str_X \str_Y \str_Z  ,
\eeq
The above bound saturates for orthogonal relative angles, $\theta_x=\theta_y=\theta_z=\frac{\pi}{2}$, which can be seen by putting $s_1(T)=s_2(T)=s_{\max}(T)$ in Eq. (\ref{relan}). For  $\str_X=\str_Y=\str_Z=1$, Eq. (\ref{corollary1eq}) and Eq. (\ref{degenerate})  reduces to the expression of the maximum expectation value of Mermin operator for non-degenerate and degenerate singular values of $T$, for projective qubit observables, respectively \cite{Siddiqui2019, PRXQuantum.2.010308}.

Now, choosing orthogonal relative angles between observables, i.e., $\theta_x=\theta_y=\theta_z=\frac{\pi}{2}$, we get the following corollary.
\begin{corollary} \label{corollary2} A sufficient condition to violate the Mermin inequality on three qubit states for unbiased observables $X,X',Y,Y',Z,Z'$, with strengths $\str_X, \str_{X'}, \str_Y,\str_{Y'},\str_Z, \str_{Z'}$  and correlation matrix T is given by
	\beq \label{corrper}
	\EuScript{M}_0^\perp:=\str_X\sqrt{(\str_Y^2\str_{Z'}^2+\str_{Y'}^2\str_Z^2)}s_1(T) + \str_{X'}\sqrt{(\str_Y^2\str_Z^2+\str_{Y'}^2\str_{Z'}^2)}s_2(T) >2 .
	\eeq
\end{corollary}
Note that Eq. (\ref{corrper}) is also a necessary criterion if the strengths of measurements are equal on every side and $s_1(T)=s_2(T)$, the optimal relative angles are orthogonal for this case (following discussion of Corollary \ref{corollary1}).

It is evident that the bound in Theorem \ref{thm3} is not invariant under the exchange of measurements, i.e., $X$ with $X'$, $Y$ with $Y'$ etc. The Mermin operator itself is not invariant under such transformations, rather gives rise to six different Mermin operators. As relative angles are invariant under such transformations, and $\sin\theta_i\geq 0$ ($i=x,y,z$), the necessary and sufficient criteria to violate any one of the six Mermin inequalities is the following:
\begin{corollary} \label{corollary3}
One of the six possible Mermin inequalities on three qubit states for unbiased observables $X,X',Y,Y',Z,Z'$, with strengths $\str_X, \str_{X'}, \str_Y,\str_{Y'},\str_Z, \str_{Z'}$  and correlation matrix $T$ is violated if only if
\begin{align} 
\label{sixpo}
\EuScript{\tilde M}_0 :=\half  [s_1(T)+s_2(T)]\tilde I_+(V) +  \half  [s_1(T)- s_2(T)] \tilde I_-(V) >2 ,
\end{align}
where $\tilde I_\pm(V) \geq 0$ is defined as
\begin{align*}
\tilde I^2_\pm(V)&=I_0+2\Big[ |I_{XY}^Z| |\cos\theta_x \cos\theta_y| + |I_{XZ}^Y||\cos\theta_x \cos\theta_z|+|I_{YZ}^X||\cos\theta_y \cos\theta_z| \nn\\&\pm \str_X \str_{X'}\sin\theta_x\Big\{I_{YZ}^0\sin^2{\theta_y}+ I_{ZY}^0\sin^2{\theta_z}+  I_1^2(1- |\cos 2\theta_y \cos 2\theta_z|) \Big\}^{\frac{1}{2}}\Big].
\end{align*}
\end{corollary} 
Comparing Eqs. (\ref{bound}) and (\ref{sixpo}) we find that $\EuScript{\tilde M}_0=\EuScript{ M}_0$, for equal strengths on each sides. 
Therefore, $\EuScript{ M}_0 >2, $ is the necessary and sufficient condition for this case, or in other words, one of the six possible Mermin inequalities violate if and only if the one considered in Eq. (\ref{form-svet-marm}) violates.

\subsection{Generalized criterion for T-state}
In this subsection, we will consider a class of mixed three qubit states whose local states are maximally mixed, i.e., $\bm l=\bm m=\bm n= \bm 0$ and additionally the bipartite correlation tensors, $\Theta= \Phi= \Omega=0$. Putting all these in Eq. (\ref{3qstate}), yields 
\begin{align}\label{T-state}
\rho_T=\frac{1}{8}[\id \otimes \id \otimes \id+ \bm \sigma^{\top} \otimes T (\bm \sigma \otimes \bm \sigma)]. 
\end{align}
These states  are therefore fully characterized by their correlation matrix $T$,  and are commonly known in literature as T-states \cite{HORODECKI1995340}. Specifically, they include GHZ-class of states [cf. \cite{Siddiqui2019}]. For this class of states, we find the upper bound of Mermin operator for most general observables, whether they are biased or unbiased observables.

Using Eq. (\ref{merm}), the form of Mermin operator for T-states can be written as
\beq \label{mop}
\EuScript{M}= |\langle \mathcal{E} \rangle_{ub}  +K|,\nn
\eeq
where  
$K:=\B_{X}(\B_{Y}\B_{Z'}+\B_{Y'}\B_{Z})+\B_{X'}(\B_{Y}\B_{Z}-\B_{Y'}\B_{Z'})$.
Noticing the similarity between $\EuScript{M}$ of unbiased measurement on arbitrary states and arbitrary measurements on T-states, we have the following theorem analogous to Theorem \ref{thm3}. 
\begin{theorem} \label{thm4}
The tight upper bound of Mermin inequality on $T$-states for arbitrary observables $X,X',Y,Y',Z,Z'$, with strengths $\str_X, \str_{X'}, \str_Y,\str_{Y'},\str_Z, \str_{Z'}$  and correlation matrix $T$ is given by 	
	\begin{align} \label{bound2}
	\EuScript{M} &\leq \EuScript{M}_{T}:= \EuScript{M}_0 +K_{\max},
	\end{align}	
	where $\EuScript{M}_0$ is same as of Eq. (\ref{bound}) and $K_{\max}$ is defined as 
\begin{align}\label{kmax}
		K_{\max}:=& (2-\str_X-\str_{X'})(2-\str_Y-\str_{Y'}) (2-\str_Z-\str_{Z'})  -r_{X}\Big( r_Y \ell_Z +\ell_Y r_Z\Big)\nn\\&  -  \ell_X\Big( r_Yr_Z +\ell_Y\ell_Z\Big)- 2r_Xr_Yr_Z,
\end{align}
where $r_i=1-\max\{\str_i,\str_{i'}\}$, $\ell_i=1-\min\{\str_i,\str_{i'}\}$, and  $i=\{X,Y,Z\}$.
	\end{theorem}
The Theorem \ref{thm4} is proved in \ref{appen2}. The Mermin inequality will be violated for $T$-state if $\EuScript{M}_{T}> 2$. Note that the optimization of the bound $\EuScript{M}_{T}$ has been subjected to the inequality in Eq. (\ref{constraint}). 

Theorem \ref{thm4} achieves a larger value for $T$-state than for general states with unbiased observables when $K_{\max}> 0$. However, for $K_{\max}= 0$, $\EuScript{M}_{T}=\EuScript{M}_0$ which indicates that unbiased observables are optimal for this case. Further, from the Eq. (\ref{kmax}), we find $K_{\max}\leq 2$ and specifically, for zero strength measurements, i.e., $K_{\max}= 2$, the values of $\EuScript{M}_0=0$ and $\EuScript{M}_{T}=2$,  implicating no violation for Mermin inequality. Then, it is interesting to ask whether there exist cases for which Mermin inequality is violated by biased observables but not by unbiased observables. In the following discussion, we answer it affirmatively.

Looking at the similarity between the upper bound of Mermin operator for unbiased measurement on arbitrary states and arbitrary measurements on T-states, we immediately have the following corollary. 
\begin{corollary} \label{corollary4}
	The tight upper bound of Mermin inequality on $T$-states, for arbitrary observables $X,X',Y,Y',Z,Z'$, with strengths $\str_X=\str_{X'}$, $\str_Y=\str_{Y'}$,  $\str_Z$=$\str_{Z'}$   and correlation matrix $T$ is given by	
	\begin{align}\label{corollary4eq}
	\EuScript{M} &\leq 2\str_X \str_Y \str_Z\sqrt{s^2_1(T)+s^2_2(T)} +2(1-\str_X)(1-\str_Y)(1-\str_Z),
	\end{align} 
where the bound is achieved for any relative angle satisfying Eq. (\ref{relan}).
\end{corollary}
Using the result of Theorem \ref{thm4} instead of the Theorem \ref{thm3}, in Eq. (\ref{corollary1eq}) of Corollary \ref{corollary1}, we get the required result, which is the sum of the right-hand side of Eq. (\ref{corollary1eq}) and $K_{\max}$.
Similar to the case of unbiased observables, the  Eq. (\ref{corollary4eq}) of arbitrary observables measured on T-states saturates when relative angles are orthogonal, i.e., when $s_1(T)=s_2(T)$.

The above corollary helps us in answering the question that we raised earlier. Let us consider an example where each observable are of the same strength $\str$, placing in Eqs. (\ref{corollary1eq}) and (\ref{corollary4eq}), we get the maximum achievable violation of the Mermin inequality for
unbiased observables, and for biased observables of equal strength  respectively,
\begin{align}\label{biased}
\EuScript{M}_{\rm unbiased} = 2\str^3 {\mathcal P},\quad{\rm and}\quad 
\EuScript{M}_{\rm biased} = 2\str^3 {\mathcal P} +2(1-\str)^3,
\end{align} 
where $\mathcal P=\sqrt{s^2_1(T)+s^2_2(T)}$. 
Comparing the above two equations, we find that if the Mermin inequality is violated by unbiased observables for a given value of $\str<1$, then it can be violated by a larger amount for the case of biased observables with the same strength. Further, there are cases where the Mermin inequality can only be violated by biased observables. Eq. (\ref{biased}) implies that a violation, i.e., $\EuScript{M}>2$, requires
\beq
\str>\str_{\rm unbiased}:=\frac{1}{\sqrt[3] {\mathcal P}},\quad
\str>\str_{\rm biased}:=\frac{-3+\sqrt{3}\sqrt{4{\mathcal P}-1}}{2({\mathcal P}-1)}. 
\eeq
Thus, we find that for any strength satisfying $\str_{\rm unbiased} \geq \str > \str_{\rm biased}$ (which is always possible if ${\mathcal P}>1$, i.e., if $M(T)>2$), the Mermin inequality can be violated only by biased observables.  
Let us summarize above results in the following corollary.

\begin{corollary} \label{corollary5} 
	The maximum violation of Mermin inequality for observables of fixed strengths $\str_X,\str_{X'},\str_Y, \str_{Y'}, \str_Z, \str_{Z'}$, 
can be larger when observables are considered biased, compared to unbiased. Moreover, there are cases where the Mermin inequality can only be violated by biased observables.  
\end{corollary}

The analogue of Corollary \ref{corollary2} for T-state, i.e., a sufficient condition to violate the Mermin inequality via general measurements of given strength, on a T-state can be written as 
\beq
\EuScript{M}_0^\perp+K_{\max} >2,\nn
\eeq
 where $\EuScript{M}_0^{\perp}$ is defined in Eq. (\ref{corrper}), while the analogue of Corollary~\ref{corollary3} is stated below.
 \begin{corollary}\label{corollary6}
One of the six possible Mermin inequalities for arbitrary observables $X,X',Y,Y',Z,Z'$, with strengths $\str_X, \str_{X'}, \str_Y,\str_{Y'},\str_Z, \str_{Z'}$ on T-states is violated if only if
	\begin{align} 
	\label{tlsixpo}
 \EuScript{\tilde M}_T:=\EuScript{\tilde M}_0 + K_{\max} > 2,
\end{align}
where $\EuScript{\tilde M}_0$ is defined in Eq. (\ref{sixpo}).
\end{corollary}
Comparing Eqs. (\ref{bound2}) and (\ref{tlsixpo}), we get $\EuScript{\tilde M}_T=\EuScript{M}_T$, for $\str_X=\str_X'$, $\str_Y=\str_Y'$ and $\str_Z=\str_Z'$. Therefore, $\EuScript{M}_T>2$, is the necessary and sufficient condition for Corollary \ref{corollary6}.
\subsection{Optimal angles for fixed strengths}
The tight upper bound of the Mermin operator obtained in Theorem \ref{thm3} \& \ref{thm4} are functions of {\it measurement strengths} and {\it relative angles} between measurement observables on each side. This dependency can be studied more in detail. Note that in the experiment, it is easier to control the measurement directions (e.g., by rotation of a polarizer) than the strength of measurement (which might come from the limitations of the device or its interaction with the environment). Therefore, we are interested to determine the optimal angles that maximize the value of the Mermin operator for a fixed set of measurement strengths. This task can always be done numerically, however, analytical results are more instructive, and thus we present some of them in this subsection.

Some of the optimal relative angles are found to be  degenerate  for the case of equal strength on each side (See Corollary \ref{corollary1} \& \ref{corollary4}), and thus those values of triple $\{\theta_x,\theta_y,\theta_z\}$ which satisfy Eq. (\ref{relan}), will be optimal. However, the degeneracy can be lifted by choosing unequal strength on just one side.  We present our first result by generalizing Corollaries \ref{corollary1} and \ref{corollary4}.
\begin{theorem}\label{fix-anglexnx}
	The tight upper bound of Mermin inequality with observables,  $X,X',Y,Y',Z,Z'$, for the cases of unbiased measurement on arbitrary states and arbitrary measurements on T-states, for strengths  $\str_X\geq \str_{X'}$, $\str_Y=\str_{Y'}$ and $\str_Z=\str_{Z'}$  and correlation matrix $T$ is given respectively by
\begin{align}
\EuScript{M} &\leq2\str_{Y}\str_{Z}\sqrt{\str_{X}^2s^2_1(T)+\str_{X'}^2 s^2_2(T)},\nn\\
\EuScript{M} &\leq2\str_{Y}\str_{Z}\sqrt{\str_{X}^2s^2_1(T)+\str_{X'}^2 s^2_2(T)} +2(1-\str_{X})(1-\str_Y)(1-\str_Z), 
\label{tbound}
\end{align}
where the bound is achieved for any relative angle satisfying
\beq 
\label{relan2m}
\sin\theta_y\sin\theta_z =\frac{ 2\str_{X}\str_{X'}s_1(T)s_2(T)}{\str^2_X s_1(T)^2+\str^2_{X'} s_2(T)^2},\:\: \mbox{and}\:\: \theta_x=\pi/2.
\eeq 
\end{theorem}
The proof of the above theorem is given in \ref{opt_angle_merm}. The results in Theorem \ref{fix-anglexnx} can be seen as a generalization of
the results found in \cite{PRXQuantum.2.010308}. We see that the results of Corollaries \ref{corollary1} \& \ref{corollary4} can be retrieved from above theorem by putting $\str_X=\str_{X'}$ and the optimal angles of Eq. (\ref{relan2m}) is also seen to satisfy the Eq. (\ref{relan}). But the optimal angles are specified uniquely, i.e., the degeneracy in angles has been lifted.
 Interestingly, the above result indicates that if two of the observers are measuring their observables with equal strengths, then the orthogonal measurement direction is the optimal choice for the third one.
 
Also, the optimal relative angles can be determined if the largest singular values of correlation matrix $T$ are equal, i.e., $s_1(T)=s_2(T)=s_{\max}(T)$, then $\EuScript{M}_0=s_{\max}(T) I_+(V)$. Finding the global maxima of $I_+(V)$ for arbitrary relative angles seems intractable, however, fixing one of the relative angles might give us some intuition. Fixing $\theta_z=\frac{\pi}{2}$, we obtained the following result. 
\begin{propos} \label{fix-smax}
	The tight upper bound of Mermin inequality for observables $X,X',Y,Y',Z,Z'$  and with two degenerate largest singular values, i.e., $s_1(T)=s_2(T) =s_{\max}(T)$ of correlation matrix $T$, is given by
	\begin{align} \label{thm4upper}
	\EuScript{M} \leq s_{\max}(T) \sqrt{I_0+2\Gamma_0}, \quad{\rm and}\quad	 \EuScript{M} \leq s_{\max}(T) \sqrt{I_0+2\Gamma_0} + K_{\max} ,
	\end{align}
	for the cases of unbiased measurements on arbitrary states and arbitrary measurements on T-states respectively, where  
	$I_0$ and $K_{\max}$ are defined respectively in 
	Eqs. (\ref{iplusminus1}) and (\ref{kmax}), and	
\begin{align*}
\Gamma_0:=\str_X \str_{X'}\sqrt{\str^2_Y \str^2_{Y'} (\str_Z^4+\str_{Z'}^4)+\str_Z^2 \str_{Z'}^2 (\str_Y^4+\str_{Y'}^4)}.
\end{align*}
  Further, the bounds are achieved for the relative angles satisfying 
	\beq 
	\tan\theta_x= \frac{\str_Z \str_{Z'} (\str_Y^2+\str_{Y'}^2)}{\str_Y \str_{Y'} (\str_Z^2-\str_{Z'}^2)},\quad\cos\theta_y={\rm sign}(\str_{X}-\str_{X'}),\quad {\rm and}\quad \theta_z=\frac{\pi}{2}.\nn
	\eeq
\end{propos}
The above proposition is proved in \ref{opt_angle_merm}. Noticing the symmetry between $\theta_y$ and $\theta_z$ in $I_+(V)$, we find that one can obtain exactly same optimization (Eq. (\ref{thm4upper})) by fixing $\theta_y=\frac{\pi}{2}$ instead of $\theta_z$. Moreover, we report that fixing $\theta_x=\frac{\pi}{2}$ will also yield the same optimization (Eq. (\ref{thm4upper})) (see \ref{opt_angle_merm} for details). Therefore, the results in Eq. (\ref{thm4upper}) seem closer to global maxima. Note that it reaches the maximum, $2\sqrt{2}s_{\max}$ when each observable is of unit strength ($\str=1$). 

\section{Generalizing Svetlichny bound}\label{Svet_unbias}

Similar to the case of the Mermin operator, we are now ready to study the Svetlichny operator for the case of unbiased observables for $\B=0$. 
The expectation value of Svetlichny operator is given by
\begin{align} \label{svetbzero}
	\langle \mathcal{E}-\mathcal{E'} \rangle_{ub} =& \str_X \bm x^{\top}  T\left[\str_Y \bm y\otimes( \str_Z \bm z +\str_{Z'} \bm z') +  \str_{Y'}\bm y'\otimes  (\str_Z \bm z -\str_{Z'} \bm z')\right]\nn\\&+\str_{X'} \bm x'^{\top}  T\left[\str_Y \bm y\otimes (\str_Z \bm z -\str_{Z'} \bm z') -  \str_{Y'}\bm y'\otimes  (\str_Z \bm z +\str_{Z'} \bm z')\right].
\end{align}

\begin{theorem}\label{thm6}
	The  tight upper bound of Svetlichny inequality on three qubit state  for unbiased observables $X,X',Y,Y',Z,Z'$, with strengths $\str_X, \str_{X'}, \str_Y,\str_{Y'},\str_Z, \str_{Z'}$  and correlation matrix $T$ is given by 
	\begin{align} \label{svetbound}
	\EuScript{S} &\leq \EuScript{S}_0:= \sum_{i=1}^{2} s_i(T) s_i(W), \nn \\ 
	&=\half  [s_1(T)+s_2(T)]J_+(W) +  \half  [s_1(T)- s_2(T)]J_-(W) ,
	\end{align}
where $W$ is a $3\times 9$ matrix defined in \ref{appen3}. And $s_1(W), s_2(W)$ are the two largest singular values of $W$ respectively, and $J_\pm(W):=s_1(W)\pm s_2(W)\geq0$, 
may be calculated using
\begin{align}\label{jplusminus1}
J^2_\pm(W)&=J_0+2\Big[J_{YZ}^X\cos\theta_x +  J_{XZ}^Y \cos\theta_y + J_{XY}^Z \cos\theta_z 
-2  \str_X \str_{X'}\Big(2I_1\cos\theta_x\cos\theta_y\cos\theta_z \nn\\ &\mp \sin\theta_x \big\{I_{YZ}^0\sin^2\theta_y + I_{ZY}^0\sin^2\theta_z+ I_1^2  (1- \cos 2\theta_y \cos 2\theta_z)\big\}^{\frac{1}{2}} \Big)\Big],
\end{align}
where $J_0=(\str_X^2+\str_{X'}^2)(\str_Y^2+\str_{Y'}^2)(\str_Z^2+\str_{Z'}^2)$, $J_{jk}^i=\str_i \str_{i'}(\str_j^2-\str_{j'}^2)(\str_k^2
	-\str_k'^2)$, and $(I_1, I_{ij}^0)$ are defined in Eq. (\ref{iplusminus1}). 
\end{theorem}
We prove the theorem in \ref{appen3}. Above theorem indicates that the Svetlichny inequality can be violated by measuring unbiased observables with given strengths and relative angles, if and only if $\EuScript{S}_0>4$. Theorem \ref{thm6} is the generalization of the special case given in Lemma \ref{theoremold2}, valid for all three qubit states. Note that the upper bound, like Mermin bound, is also invariant under local unitary transformations on every side, since such transformations leave measurement strengths, relative angles, and singular values invariant. For equal strength on each side yields the following bound.
\begin{corollary} \label{corollary7}
	The tight upper bound of Svetlichny inequality on three qubit states for unbiased observables $X,X',Y,Y',Z,Z'$, with strengths $\str_X=\str_{X'}$, $\str_Y=\str_{Y'}$,  $\str_Z$=$\str_{Z'}$ and correlation matrix $T$ is given by	
	\beq \label{corollary7eq}
	\EuScript{S}\leq 2\sqrt{2} \str_X\str_Y\str_Z \sqrt{s^2_1(T)+s^2_2(T)},
	\eeq
The above bound is achieved for any relative angle satisfying
	\begin{align} \label{relan7}
    \cos \theta_y \cos \theta_z	 = \frac{s^2_1(T)-s^2_2(T)}{s^2_1(T)+s^2_2(T)} \quad {\rm and}\quad \theta_x=\frac{\pi}{2},\quad \nn\\
{\rm or,}\quad \sin\theta_x=\frac{2s_1(T)s_2(T)}{s^2_1(T)+s^2_2(T)}, \quad {\rm and}\quad \cos\theta_y\cos\theta_z=0.
	\end{align}
\end{corollary}
The proof of corollary is given in \ref{appen3}.   
The new bound obtained in Corollary \ref{corollary7} is a substantial generalization of the Lemma \ref{theoremold2} for unbiased observables.
For two degenerate largest singular values of $T$, say $s_{\max}(T)$, Eq. (\ref{corollary7eq}) becomes
\beq \label{degeneratesvet}
\EuScript{S}  \leq 4 s_{\max}(T) \str_X \str_Y \str_Z.
\eeq
The above bound saturates, for orthogonal relative angles, $\theta_i=\frac{\pi}{2}$, which can been seen by putting $s_1(T)=s_2(T)=s_{\max}(T)$ in Eq. (\ref{relan7}). 
For  $\str_X=\str_Y=\str_Z=1$, 
Corollary \ref{corollary7} reduces to the following upper bound of Svetlichny inequality with sharp observables, 
\begin{align} \label{svet}
\EuScript{S} \leq S(T):= 2\sqrt{2} \sqrt{s^2_1(T)+s^2_2(T)}.
\end{align}
Eq. (\ref{svet}) is a generalization of a special case represented in Lemma \ref{theoremold2}, for sharp observables. When $s_1(T)=s_2(T)=s_{\max}(T)$, we retrieve the Lemma \ref{theoremold2} \cite{PhysRevA.96.042323}.

It follows from Theorem \ref{thm6} that the sufficient condition to violate Svetlichny inequality for fixed values of relative angles only depends on the measurement strengths. Specifically,  by choosing orthogonal relative angles, $\theta_x=\theta_y=\theta_z=\frac{\pi}{2}$, we get the following criteria. 
\begin{corollary} \label{corollary8} A sufficient condition to violate the Svetlichny inequality on three qubit states for unbiased observables $X,X',Y,Y',Z,Z'$, with strengths $\str_X, \str_{X'}, \str_Y,\str_{Y'},\str_Z, \str_{Z'}$  and correlation matrix T is given by
\beq \label{corrper8}
\EuScript{S}_0^\perp:=\half  (j_++j_-)s_1(T) + \half  (j_+-j_-)s_2(T) >4 ,
\eeq
where
$j_\pm:= \Big\{J_0\pm 4 \str_X \str_{X'}
\sqrt{(\str_Y^2\str_{Z'}^2+\str_{Y'}^2\str_Z^2)(\str_Y^2\str_Z^2+\str_{Y'}^2\str_{Z'}^2)}\Big\}^{\frac{1}{2}}$. 	
\end{corollary}
Eq. (\ref{corrper8}) is also a necessary condition for the case of equal strengths on each side and $s_1(T)=s_2(T)$, the optimal relative angles are orthogonal for this case (follow discussion of Corollary \ref{corollary7}).

Following similar reasoning like Mermin inequality, the Svetlichny operator is also not invariant under the exchange of measurement operators on each side, i.e., $X\to X'$, $Y\to Y'$ etc, rather, it transforms between six different versions of Svetlichny operators. Therefore, Theorem \ref{thm6} is not also invariant under such transformations. As the relative angles are invariant under such transformations, and $\sin \theta_i\geq 0$, the necessary and sufficient condition to violate any one of six Svetlichny inequalities is stated in the following corollary.
\begin{corollary} \label{corollary9}
	One of the six possible Svetlichny inequalities on three qubit states for unbiased observables $X,X',Y,Y',Z,Z'$, with strengths $\str_X, \str_{X'}, \str_Y,\str_{Y'},\str_Z, \str_{Z'}$  and correlation matrix $T$ is violated if only if
\begin{align} \label{sixposv}
	\EuScript{\tilde S}_0 :=\half  [s_1(T)+s_2(T)]\tilde J_+(W) +  \half  [s_1(T)- s_2(T)] \tilde J_-(W) >4 ,
\end{align}
with $\tilde J_\pm(W) \geq 0$ defined via 
\begin{align}
\tilde J^2_\pm(W)&=J_0+2\Big[|J_{YZ}^X||\cos\theta_x| + |J_{XZ}^Y| |\cos\theta_y| + |J_{XY}^Z| |\cos\theta_z|
-2  \str_X \str_{X'}\Big(2I_1|\cos\theta_x\cos\theta_y\nn\\ &\times \cos\theta_z| \mp \sin\theta_x \big\{I_{YZ}^0\sin^2\theta_y +I_{ZY}^0\sin^2\theta_z+ I_1^2 (1- |\cos 2\theta_y \cos 2\theta_z|)\big\}^{\frac{1}{2}} \Big)\Big].\nn
\end{align}
\end{corollary}
Comparing Eqs. (\ref{svetbound}) and (\ref{sixposv}), we find that $\EuScript{\tilde S}_0=\EuScript{ S}_0$, for equal strengths on each side.  Therefore, $\EuScript{ S}_0 >2, $ is the necessary and sufficient condition for this case, or in other words, one of the six possible Svetlichny  inequalities violate if and only if the one considered in Eq. ({\ref{form-svet-marm}}) violates.

\subsection{Generalized criterion for T-state}
Now, we obtain the upper bound of Svetlichny operator for T-states which is valid for arbitrary observables, whether biased or unbiased. The form of Svetlichny operator for T-states is given by
\beq
\EuScript{S}=  |\langle \mathcal{E}-\mathcal{E'} \rangle_{ub}+L|\nn,
\label{sop}
\eeq
where 
$L:=(\B_{X}\B_{Y}-\B_{X'}\B_{Y'})(\B_{Z}+\B_{Z'})+(\B_{X}\B_{Y'}+\B_{X'}\B_{Y})(\B_{Z}-\B_{Z'})$.
Noticing the similarity between $\EuScript{S}$ of unbiased measurement on arbitrary states and arbitrary measurements on T-states, we have the following theorem, equivalent to Theorem \ref{thm6}.
\begin{theorem} \label{thm7}
	The tight upper bound of Svetlichny inequality on $T$-states for arbitrary observables $X,X',Y,Y',Z,Z'$, with strengths $\str_X, \str_{X'}, \str_Y,\str_{Y'},\str_Z, \str_{Z'}$  and correlation matrix $T$ is given by 
\begin{align} \label{bound5}
	\EuScript{S} &\leq \EuScript{S}_{T}:= \EuScript{S}_0 +L_{\max},
\end{align}	
where $\EuScript{S}_0$ is same as of Eq. (\ref{svetbound}) and $L_{\max}$ is defined as
\begin{align}
L_{\max}:=& \left[(1-\str_X)(1-\str_{Y'})+(1-\str_{X'})(1-\str_Y)\right]|2-\str_Z-\str_{Z'}|\nn\\&+\Big[(1-\str_X)(1-\str_Y)-(1-\str_{X'})(1-\str_{Y'})\Big]|\str_{Z'}-\str_Z|.
\label{lmax}
\end{align}
\end{theorem}
 The proof of Theorem \ref{thm7} is given in \ref{appen4}. The Svetlichny inequality will be violated  for given strengths and relative angles on T-states, if and only if $\EuScript{S}_{T}>4$. Note that we achieve the expression of $L_{\max}$ by optimizing the Svetlichny expression subjected to the constraint in Eq. (\ref{constraint}).
  
 As $L_{\max}\geq 0$, it can be concluded from Theorem \ref{thm6} \& \ref{thm7} that the Svetlichny operator can achieve a larger value for $T$-states compared to unbiased observables for all three qubit states. However, for $L_{\max}=0$, i.e., all strengths equal to unity, $\EuScript{S}_{T}=\EuScript{S}_{0}$, implying that the unbiased observables are optimal in this case. Further, from Eq. (\ref{lmax}), we find $L_{\max}\leq 4$ and specifically, for zero strength measurements, i.e., $L_{\max}= 4$, the values of $\EuScript{S}_0=0$ and $\EuScript{S}_{T}=4$,  implicating no violation for Svetlichny inequality.  Again, this fact raises a similar question: does there exist instances where Svetlichny inequality will be violated by biased observables but not by unbiased ones? We answer it affirmatively in the following discussions. 

Noticing the similarity between Theorem \ref{thm6} \& \ref{thm7}, one can derive the following corollary.
\begin{corollary} \label{corollary10}
	The tight upper bound of Svetlichny inequality on $T$-states, for arbitrary observables $X,X',Y,Y',Z,Z'$, with strengths $\str_X=\str_{X'}$, $\str_Y=\str_{Y'}$,  $\str_Z$=$\str_{Z'}$ and correlation matrix $T$ is given by	
	\begin{align}\label{corollary10eq}
	\EuScript{S} \leq  2\sqrt{2} \str_X\str_Y\str_Z 
	\sqrt{s^2_1(T)+s^2_2(T)} + 4(1-\str_{X})(1-\str_Y)(1-\str_Z).
	\end{align} 
	The above bound is achieved for any relative angle satisfying Eq. (\ref{relan7}).
\end{corollary}
Using the result of Theorem \ref{thm7} instead of that Theorem \ref{thm6}, in Eq. (\ref{corollary7eq}) of Corollary \ref{corollary7}, we get the required result, which is sum of the right hand side of Eq. (\ref{corollary7eq}) and $L_{\max}$.
Similar to the case of unbiased observables, the  Eq. (\ref{corollary10eq}) of arbitrary observables measured on T-states saturates when relative angles are orthogonal,  $\theta_x=\theta_y=\theta_z=\frac{\pi}{2}$, i.e., when $s_1(T)=s_2(T)$.

The above corollary helps us in answering the question that we raised earlier. Let us consider an example where each observable are of the same strength $\str$, placing in Eqs. (\ref{corollary7eq}) and  (\ref{corollary10eq}), we get the maximum achievable violation of the Svetlichny inequality for unbiased observables, and for biased observables of equal strength respectively,
\begin{align}\label{biasedsvet}
\EuScript{S}_{\rm unbiased} = 2\sqrt{2}\str^3 {\mathcal P},\:\:
\mbox{and}\:\:
\EuScript{S}_{\rm biased} = 2\sqrt{2}\str^3 {\mathcal P} +4(1-\str)^3.
\end{align}
where $\mathcal P=\sqrt{s^2_1(T)+s^2_2(T)}$. 
Comparing the above two equations, we find that if the Svetlichny inequality is violated by unbiased observables for a given value of $\str<1$, then, it can be violated by a larger amount for the case of biased observables with the same strength. Further, there are cases where the Svetlichny inequality can only be violated by biased observables. Eq. (\ref{biasedsvet}) implies that a violation $\EuScript{S}>4$ requires
\beq
\str>\str_{\rm unbiased}:=\sqrt[3] {\frac{\sqrt{2}}{\mathcal P}},\quad 
\str>\str_{\rm biased}:=\frac{-3+\sqrt{3}\sqrt{2\sqrt{2}{\mathcal P}-1}}{\sqrt{2}({\mathcal P}-\sqrt{2})}.
\eeq
Thus, for $\str_{\rm unbiased} \geq \str > \str_{\rm biased}$ (which is always possible if ${\mathcal P}> \sqrt{2}$, i.e., if $S(T)>4$), 
we see that there exist cases where the Svetlichny inequality can only be violated by biased observables.  Let us summarize above result in a Corollary \ref{corollary11}.
\begin{corollary} \label{corollary11} 
The maximum violation of Svetlichny inequality for observables of fixed strengths $\str_X,\str_{X'},\str_Y, \str_{Y'}, \str_Z, \str_{Z'}$, can be larger when observables are biased as compared to unbiased. Moreover, there are cases where the Svetlichny inequality can only be violated by biased observables.  
\end{corollary}

The analogue of Corollary \ref{corollary8} for T-state, i.e., a sufficient condition to violate the Svetlichny inequality via general measurements of given strength, on T-states can be written as 
\beq
\EuScript{S}_0^\perp+L_{\max} >4,\nn
\eeq
where $\EuScript{S}_0^{\perp}$ is defined in Eq. (\ref{corrper8}), while the analogue of Corollary~\ref{corollary9} is stated below.
\begin{corollary}
	One of the six possible Svetlichny inequalities for arbitrary observables $X,X',Y,Y',Z,Z'$, with strengths $\str_X, \str_{X'}, \str_Y,\str_{Y'},\str_Z, \str_{Z'}$ on T-states is violated if only if
	\begin{align} \label{svetsix}
		\EuScript{\tilde S}_T:=\EuScript{\tilde S}_0 + L_{\max} > 4,
	\end{align}
	where $\EuScript{\tilde S}_0$ is defined in Eq. (\ref{sixposv}).
\end{corollary} 
Comparing Eqs. (\ref{bound5}) and (\ref{svetsix}), we get $\EuScript{\tilde S}_T=\EuScript{ S}_T$, for $\str_X=\str_X'$, $\str_Y=\str_Y'$ and $\str_Z=\str_Z'$. Therefore, $\EuScript{ S}_T>4$, is the necessary and sufficient condition for  
above corollary.
\subsection{Optimal angles for fixed strengths}
The tight upper bound of the Svetlichny operator, established in Theorem \ref{thm6} \& \ref{thm7}, are functions of both the measurement strengths and the relative angles of local measurements. We are aware that the dependence on angles can be easily controlled in experiments, but not the strengths. Hence, we intend to determine the optimal angles that maximizes the value of Svetlichny operator when the strengths of local measurements are fixed. Though it can be achieved numerically, the analytical analysis are more instructive. We present some of the results in the following. 
 
Some of the optimal relative angles are found to be  degenerate for the case of equal strength on each side (See Corollary \ref{corollary7} \& \ref{corollary10}), and thus, those values of triple $\{\theta_x,\theta_y,\theta_z\}$ which satisfy Eq. (\ref{relan7}), will be optimal. However, the degeneracy can be lifted by choosing unequal strength on just one side.  We present our first result which generalizes the Corollaries \ref{corollary7} and \ref{corollary10}.
\begin{theorem}\label{svet_maxeq}
	The tight upper bound of Svetlichny inequality for observables,  $X,X',Y,Y',Z,Z'$, for the cases of unbiased measurement on arbitrary state and arbitrary measurements on T-states, for strengths  $\str_X\geq \str_{X'}$, $\str_Y=\str_{Y'}$ and $\str_Z=\str_{Z'}$  and correlation matrix $T$ is given respectively by
	\begin{align}
	\EuScript{S} \leq \EuScript{S}_0^{*},\:\:\mbox{and}\:\: 
	\EuScript{S} \leq  \EuScript{ S}_0^{*} +2(2-\str_{X}-\str_{X'})(1-\str_Y)(1-\str_Z), \label{tbound2}
	\end{align}
where values of $\EuScript{ S}_0^{*}$ are given for different choices of angles 
\begin{align*}
\EuScript{ S}_0^{*}=\left\{ 
\begin{array}{ll}
2\str_{Y}\str_{Z}\Big[\str_{X}s_1(T)+\str_{X'}s_2(T)\Big]	 , & \quad  \theta_{y} =\theta_z=\frac{\pi}{2}, \:\:\theta_x=\frac{\pi}{2}, \\
2 \str_{Y}\str_{Z}\sqrt{\str^2_X+\str^2_{X'}}\Big[s_1^2(T)+s_2^2(T)\Big]^\frac{1}{2} , & \quad 
\sin\theta_y\sin\theta_z=\frac{2s_1(T)s_2(T)}{s^2_1(T)+s^2_2(T)},\:\: \theta_x=\frac{\pi}{2} ,\\
2\sqrt{2} \str_{Y}\str_{Z}s_{\max}(T)\sqrt{\str^2_X+\str^2_{X'}}, & \quad 
\sin\theta_y\sin\theta_z=\frac{|\str^2_X-\str^2_{X'}|}{\str^2_X+\str^2_{X'}},\:\: \theta_x=0,
\end{array}
\right.
\end{align*}
where  
$s_{\max}(T)$ is the largest singular value with degeneracy two. 
\end{theorem}
The proof of above theorem is given in \ref{opt_angle_svet}. These results lift the degeneracies in relative angles from Corollaries \ref{corollary7} \& \ref{corollary10}, i.e., angles are specified uniquely for $\str_X\geq \str_{X'}$. Interestingly, Theorem \ref{svet_maxeq} indicates that if two of the parties are measuring their observables with equal strengths, then orthogonal as well as parallel measurement directions are optimal for the third one, provided it satisfies the constraints on $\theta_y$ and $\theta_z$. It also substantially improve the previous results surmised in Lemma \ref{theoremold2}. Note that one retrieves the corollaries for $\str_X=\str_{X'}$.

The optimal angles which will optimize the Svetlichny operator can also be determined when $s_1(T)=s_2(T)=s_{\max}(T)$. Then, $\EuScript{S}_0=s_{\max}(T)J_+(W)$. Finding global maxima of $J_+(W)$ for arbitrary angles seems challenging. However, fixing one of the angles might serve our purpose. We obtain the following result after fixing $\sin\theta_x=0$.
\begin{propos} \label{fix-smaxsv}
	The tight upper bounds of Svetlichny inequality for observables $X,X',Y,Y',Z,Z'$ and  with two degenerate largest singular values, i.e., $s_1(T)=s_2(T) =s_{\max}(T)$ of correlation matrix $T$, is given by
	\begin{align} \label{thm4uppersv}
	\EuScript{S} \leq s_{\max}(T) 
\sqrt{J_0+2\Gamma_1}, \qquad	 \EuScript{S} \leq s_{\max}(T) 
\sqrt{J_0+2\Gamma_1} + L_{\max} ,
	\end{align}
	for the cases of unbiased measurements on arbitrary states and arbitrary measurements on T-states respectively with $\cos\theta_x={\rm sign}([\str_Y -\str_{Y'}][\str_Z-\str_{Z'}])$, where 
\begin{align*}
\Gamma_1=&|J_{YZ}^X|+|J_{XZ}^Y|+|J_{XY}^Z|+4  \str_X \str_{X'}\str_Y \str_{Y'} \str_Z \str_{Z'},
\end{align*}
with $(J_0, J_{ij}^k)$ and $L_{\max}$ are defined respectively in Eqs. (\ref{jplusminus1})  and (\ref{lmax}).
	Further, the bounds are achieved for the relative angles satisfying
	\beq 
	\cos\theta_y={\rm sign}\left([\str_{X}-\str_{X'}][\str_{Z}-\str_{Z'}]\right),\quad \cos\theta_z={\rm sign}\left([\str_{X}-\str_{X'}][\str_{Y}-\str_{Y'}]\right), \quad \nn
	\eeq
\end{propos}

The proof of the above proposition is given in \ref{opt_angle_svet}. 
We find that $s_{\max}(T)\sqrt{J_0+2\Gamma_1}$ reaches its maximum value $4s_{\max}(T)$ for the measurements with unit strengths. Proposition \ref{fix-smaxsv} can be generalized further for arbitrary $\theta_x$, which we have left for future exploration.

\section{Conclusions and future directions}\label{concl}
We have studied the Mermin and Svetlichny inequalities with dichotomic non-projective measurement observables for general three qubit states. Our analysis is also relevant for the practical scenarios where analysis of Bell inequalities with non-projective measurement are either preferable or inevitable. We find the quantum upper bound of Mermin as well as Svetlichny operators for unbiased observables on arbitrary three qubit states, and for arbitrary measurements on $T$-states. Also, we find the necessary and sufficient conditions to violate these inequalities by arbitrary three qubit states for measurements with fixed strengths and relative angles for all local parties. Also, we determine the optimal angles to violate these inequalities for fixed as well as arbitrary measurement strengths. Our results significantly generalize the previous extant bounds \cite{Siddiqui2019, PRXQuantum.2.010308,PhysRevA.96.042323} in the following aspects:
\begin{itemize}
\item Theorem \ref{thm3} substantially generalizes the upper bound of Mermin operator with unbiased observables for three qubit states, previously obtained in Ref. \cite{Siddiqui2019, PRXQuantum.2.010308}.
\item Theorem \ref{thm4} represents a new set of bounds for Mermin operators for arbitrary measurement observables for $T$-states.
\item Theorem \ref{thm6} improves hugely the earlier bounds of Svetlichny operator found in Ref. \cite{PhysRevA.96.042323}. In fact, the extant bound was tight only in the cases where the largest singular value of correlation matrix $T$ has degeneracy two. 
\item Theorem \ref{thm7} states a new set of bound of Svetlichny inequality valid for arbitrary measurements on $T$-states.
\item We also obtained the necessary and sufficient criteria to violate the Mermin and Svetlichny inequality as well as their other five variants.
\end{itemize}
Then, we consider some special cases of Theorems \ref{thm3} and \ref{thm6} and find the optimal relative angles which will lead us to the maximal value of Mermin and Svetlichny operators for arbitrary three qubit states. We believe that these bounds might help us to find the one-sided monogamy relations, similar to Bell-CHSH inequalities found in Refs. \cite{PhysRevA.104.L060201,2022arXiv220301837G}. These results can be useful to the concept of recycling Bell nonlocality for multipartite settings \cite{PhysRevA.103.032216,2019QuIP...18...42S}, which we will pursue in our future work. Our analysis can readily be applied to the  other multipartite Bell inequalities, like, MABK inequalities \cite{PhysRevLett.65.1838,PhysRevA.46.5375,Belinskii:1993} and other facet inequalities \cite{2003PhLA..317..165S,2017PhLA..381.3928D}. It would be of great interest to find a full generalization of our work for the case of  arbitrary measurements and arbitrary three qubit states. Several experiments have been performed earlier to test the violation of the Mermin and Svetlinchy inequalities \cite{PMID:10676953,Lavoie_2009, Swain2019}. We expect a similar experimental work on our results in the future.

{\flushleft{\it Acknowledgements:}
 MAS acknowledges the National Key R$\&$D  Program  of  China,  Grant  No. 2018YFA0306703. SS acknowledges the financial support through the {\v S}tefan Schwarz stipend from Slovak Academy of Sciences, 
Bratislava. SS also acknowledges the financial support through the project OPTIQUTE (APVV-18-0518), HOQIP (VEGA 2/0161/19) and DESCOM (VEGA-2/0183/21).}


\appendix 
\section{Proof of Theorem \ref{thm3} and Corollary \ref{corollary1}}\label{appen}
The proof of Theorem \ref{thm3} is as follows. 
\begin{proof} For two-valued qubit observables $X,X',Y,Y',Z,Z'$, represented by Eq. (\ref{observable}), we define the unit vectors
\beq \label{xdefn}
\bm x_1 = \frac{\bm x+ \bm x'}{|\bm x+\bm x'|},\quad \bm x_2 = \frac{\bm x- \bm x'}{|\bm x - \bm x'|}, \quad \bm x_3=\bm x_1\times \bm x_2, \nn
\eeq
\beq \label{ydefn}
\bm y_1 = \frac{\bm y+ \bm y'}{|\bm y + \bm y'|}, \quad \bm y_2 = \frac{\bm y- \bm y'}{|\bm y - \bm y'|},\quad  \bm y_3=\bm y_1\times \bm y_2,  \nn
\eeq
\beq \label{zdefn}
\bm z_1 = \frac{\bm z+ \bm z'}{|\bm z + \bm z'|},\quad \bm z_2 = \frac{\bm z- \bm z'}{|\bm z - \bm z'|}, \quad \bm z_3=\bm z_1\times \bm z_2.\nn
\eeq
Then we can write the original measurement vectors as 
\beq \label{x1pmx2}
\bm x =\cos\frac{\theta_x}{2} \bm x_1+\sin \frac{\theta_x}{2} \bm x_2, \quad \bm x' =\cos\frac{\theta_x}{2} \bm x_1-\sin \frac{\theta_x}{2} \bm x_2, \nn
\eeq
\beq \label{y1pmy2}
\bm y =\cos\frac{\theta_y}{2} \bm y_1+\sin \frac{\theta_y}{2} \bm y_2, \quad \bm y' =\cos\frac{\theta_y}{2} \bm y_1-\sin \frac{\theta_y}{2} \bm y_2 ,\nn
\eeq
\beq \label{z1pmz2}
\bm z =\cos\frac{\theta_z}{2} \bm z_1+\sin \frac{\theta_z}{2} \bm z_2, \quad \bm z' =\cos\frac{\theta_z}{2} \bm z_1-\sin \frac{\theta_z}{2} \bm z_2 ,\nn
\eeq
where $\cos \theta_x=\bm x\cdot\bm x'$, $\cos \theta_y=\bm y\cdot\bm y'$ and  $\cos \theta_z=\bm z\cdot\bm z'$ with $0\leq\theta_i\leq\pi$ ($i=x,y,z$).
Therefore, Eq. (\ref{mermbzero}) can be simplified to
\begin{align}\label{m-matrix}
\langle \mathcal{E} \rangle =\sum_{ijk}V_{i(jk)} \bm x_i^\top T (\bm y_j\otimes \bm z_k)={\rm Tr}[VM^\top],
\end{align}
where the matrix $V$ is defined as
      \begin{align*}
	V=\begin{pmatrix}
	V_1& \bm 0^{\top}&V_2&  (0) \\
	\bm 0&0&\bm 0&\bm 0
	\end{pmatrix},\:\:\mbox{where}\:\: 
	V_1=\begin{pmatrix}
	A\cos\frac{\theta_x}{2}\cos\frac{\theta_y}{2}\cos\frac{\theta_z}{2}&  B\cos\frac{\theta_x}{2}\cos\frac{\theta_y}{2}\sin\frac{\theta_z}{2} \\
	C\sin\frac{\theta_x}{2}\cos\frac{\theta_y}{2}\cos\frac{\theta_z}{2}& -D\sin\frac{\theta_x}{2}\cos\frac{\theta_y}{2}\sin\frac{\theta_z}{2}   
	\end{pmatrix},
	\end{align*}
with $\bm 0$ is null vector, $(0)$ denotes $2\times 4$ null matrix and
       \begin{align*}
	V_2=\begin{pmatrix}
	D \cos\frac{\theta_x}{2}\sin\frac{\theta_y}{2}\cos\frac{\theta_z}{2}
	& -C \cos\frac{\theta_x}{2}\sin\frac{\theta_y}{2}\sin\frac{\theta_z}{2}\\ 
	-B \sin\frac{\theta_x}{2}\sin\frac{\theta_y}{2}\cos\frac{\theta_z}{2}& -A \sin\frac{\theta_x}{2}\sin\frac{\theta_y}{2}\sin\frac{\theta_z}{2}
	\end{pmatrix},
	\end{align*}
	where $A, B, C,$ and $D$ are given by 
	\begin{align}\label{abcd}
	A&= \str_X \str_Y \str_{Z'} + \str_X \str_{Y'} \str_Z+ \str_{X'} \str_Y \str_{Z}-\str_{X'}  \str_{Y'} \str_{Z'} \nn \\
	B&= -\str_X \str_Y \str_{Z'} + \str_X \str_{Y'} \str_Z+ \str_{X'} \str_Y \str_{Z}+\str_{X'}  \str_{Y'} \str_{Z'} \nn \\
	C&= \str_X \str_Y \str_{Z'} + \str_X \str_{Y'} \str_Z- \str_{X'} \str_Y \str_{Z}+\str_{X'}  \str_{Y'} \str_{Z'} \nn \\
		D&= \str_X \str_Y \str_{Z'} - \str_X \str_{Y'} \str_Z+ \str_{X'} \str_Y \str_{Z}+\str_{X'}  \str_{Y'} \str_{Z'},
	\end{align}
and $M$ is the $3\times9$ matrix with coefficients
\beq
M_{i(jk)} := \bm x_i^\top T \bm y_j\otimes \bm z_k.
\eeq
The sub-matrices $V_1$ and $V_2$ contains the information of local measurement strengths and relative angles, whereas $M$ contains the global information of three body spin correlations. Notice that the entries in the $3\times 9$ matrix, $M$ depends on the entries of the correlation matrix $T$. However, the evaluation of the expression in Eq. (\ref{m-matrix}) does not depend on the way one chooses the entries of $T$, as it will also shuffle the entries of $V$ accordingly. The further evaluation of Eq. (\ref{m-matrix}) depends on the following lemma by von Neumann. Let us first state the Lemma 
\begin{lemma}\label{lemma1}  
Horn and Johnson \cite{horn} Let $A$ and $B$ are m × n rectangular matrices, and 
	$s_1(A)\geq s_2(A) \dots \geq s_r(A)$ and $s_1(B)\geq s_2(B) \dots \geq s_r(B)$ denote the non-increasingly ordered singular values of A and B, respectively. Then the following relation holds 
	\begin{align}
	|{\rm Tr}AB^{\top}|\leq \sum_{i=1}^{r}s_i(A)s_i(B),
	\end{align} 
	where $r=\min\{m,n\}$.
\end{lemma}
The above lemma can be used to prove our theorem. First, we notice that one of the singular values of the matrix $V$ is zero, i.e., $s_3(V)=0$. Therefore, the Eq. (\ref{m-matrix}) simplifies to the first equation of Theorem \ref{thm3}, i.e., 
\begin{align}\label{fnlbound}
\EuScript{M}\leq   \sum_{i=1}^{2} s_i(V)s_i(T).
\end{align}
Now, we will check whether the bound in Eq. (\ref{fnlbound}) can be achieved. 
\begin{proof}
Because of the choice of unit vectors, one can see that $\bm y_j=R_2 \bm x_j$ and $\bm z_k=R_3 \bm x_k$, where $R$'s are some rotation in 3D. Therefore, one reaches to 
\begin{align*}
M_{ijk}=\bm x_i^\top T (R_2\otimes R_3) (\bm x_j \otimes \bm x_k),
\end{align*}
which immediately shows $s_i(M)=s_i(T)$ in Eq. (\ref{fnlbound}). 

Next, arbitrary orthogonal rotations $O_1$, $O_2$ and $O_3$ respectively on $\{\bm x_i\}$, $\{\bm y_j\}$ and $\{\bm z_k\}$ will keep $V$ invariant, while $M$ changes to 
\begin{align*}
M'_{ijk}=\bm x_i^\top O_1^\top T (O_2 R_2\otimes O_3 R_3)(\bm x_j \otimes \bm x_k).
\end{align*} 
Further, we find that 
\begin{align*}
{\rm Tr}[VM'^\top]={\rm Tr}[V O_1^\top T (O_2 R_2\otimes O_3 R_3)].
\end{align*}
Let $V=P_1 V_{d} P_2^\top$ and $T=Q_1 T_{d} Q_2^\top$ be the singular value decompositions of $V$ and $T$ for some orthogonal matrices $P$'s and $Q$'s, where $[V_d]_{ii}=s_i(V)$ and $[T_d]_{ii}=s_i(T)$. Then, 
\begin{align*}
{\rm Tr}[VM'^\top]&={\rm Tr}[V_d P_2^\top O_1^\top Q_1 T_d Q_2^\top (O_2 R_2\otimes O_3 R_3)P_1],\\
&={\rm Tr}[V_d T_d]\\
&=s_1(V)s_1(T)+s_2(V)s_2(T),
\end{align*} 
where we choose
 \beq\label{satcon}
  O_1^\top=P_2 Q_1^\top  \qquad \mbox{and} \qquad O_2 R_2\otimes O_3 R_3=Q_2 P_1^\top
 \eeq
 in the second line and $s_3(V)=0$ in third line. This completes the proof that the bound will saturate provided that the appropriate local transformations defined in Eq. (\ref{satcon}) are applied. 
\end{proof}

To obtain the alternate expression in Theorem \ref{thm3}, we notice that $\EuScript{M}_0$ can be rewritten as 
\begin{align*}
\EuScript{M}_0=\half  [s_1(T)+s_2(T)]I_+(V) +  \half  [s_1(T)- s_2(T)]I_-(V),
\end{align*}
where $I_\pm(V)=s_1(V)\pm s_2(V)=\sqrt{v_+}\pm\sqrt{v_-}$, with $v_\pm$ are the eigenvalues of $V^\top V$. Using the identities $v_++v_-={\rm Tr}[V^\top V]$ and $v_+v_-={\rm det} (V^\top V)={\rm det} (V)^2$, one reaches 
\begin{align*}
I_\pm(V)^2={\rm Tr}[V^\top V]\pm 2|{\rm det} (V)|.
\end{align*}
Now explicit calculations of terms in the above equation will give us the expression of $I_\pm(V)$.
\end{proof}

Following is the proof of Corollary \ref{corollary1}.
\begin{proof}
To obtain Eq. (\ref{corollary1eq}),  we first calculate the eigenvalues $\textsl{v}_\pm$ of $V^\top V$ for $\str_X=\str_{X'}$, $\str_Y=\str_{Y'}$ and $\str_Z=\str_{Z'}$,
\beq \label{vpm}
\textsl{v}_\pm= 2 \str_X^2 \str_Y^2 \str_Z^2 \left(1\pm \sqrt{1-\sin ^2 \theta_x \left(1- \cos ^2 \theta_y \cos ^2 \theta_z \right)}\right).
\eeq
Applying Cauchy-Schwarz inequality to Eq. (\ref{fnlbound}), we get
\begin{align} \label{cor1prf}
\EuScript{M} &\leq \left(\sum_{i=1}^{2} s^2_i(V)\right)^{1/2} \left(\sum_{i=1}^{2} s^2_i(T)\right)^{1/2}, \nn\\
&= \sqrt{\textsl{v}_++\textsl{v}_-}\sqrt{s^2_1(T)+s^2_2(T)},\nn\\
&=2\str_X\str_Y\str_Z\sqrt{s^2_1(T)+s^2_2(T)},
\end{align}
where $\textsl{v}_+ =s^2_1(V), \textsl{v}_-=s^2_2(V)$ and equality holds if and only if  $\textsl{v}_+/\textsl{v}_-=s^2_1(T)/s^2_2(T)$.
Using Eq. (\ref{vpm}) we get the  equality condition
\beq \label{relan2}
\sin \theta_x \sqrt{1-\cos^2 \theta_y \cos^2 \theta_z} = \frac{2s_1(T)s_2(T)}{s^2_1(T)+s^2_2(T)}  .
\eeq
Hence the results.
\end{proof}
\section{Proof of Theorem \ref{thm4}}\label{appen2}
\begin{proof}
 Following Eq. (\ref{constraint}), we see that $|\B|\leq\bar\str:= 1-\str$. We define, 
		$\B_X=\alpha\bar\str_X$, $\B_{X'}=\alpha'\bar\str_{X'}$, $\B_Y=\beta\bar\str_Y$, $\B_{Y'}=\beta'\bar\str_{Y'}$, $\B_Z=\gamma\bar\str_Z$, $\B_{Z'}=\gamma'\bar\str_{Z'}$, for suitable choices of $\alpha_,\alpha',\beta,\beta',\gamma,\gamma'=\pm1$, and $\delta=\beta'/\beta$, $\nu=\gamma'/\gamma$.
		Substituting back in $K$, we get
		\begin{align}
		|K|\leq&\max_{\alpha,\alpha',\beta,\beta',\gamma,\gamma'=\pm1} |\alpha\bar\str_X(\beta\gamma'\bar\str_Y\bar\str_{Z'}+\beta'\gamma\bar\str_{Y'}\bar\str_Z) + \alpha'\bar\str_{X'}(\beta\gamma\bar\str_Y\bar\str_Z-\beta'\gamma'\bar\str_{Y'}\bar\str_{Z'})| \nn\\
		=& \max_{\delta,\nu=\pm1}\, \bar\str_{X}|\bar\str_Y\bar\str_{Z'}+\frac{\delta}{\nu}\bar\str_{Y'}\bar\str_Z| + \bar\str_{X'}|\bar\str_Y\bar\str_Z-\delta\nu\bar\str_{Y'}\bar\str_{Z'}| \nn\\
		=&\max\{\bar\str_X,\bar\str_{X'}\}(\bar\str_Y\bar\str_{Z'}+\bar\str_{Y'}\bar\str_Z) + \min\{\bar\str_X,\bar\str_{X'}\}|\bar\str_Y\bar\str_Z-\bar\str_{Y'}\bar\str_{Z'}|\nn\\
		=& (\bar\str_X+\bar\str_{X'})(\bar\str_{Y}+\bar\str_{Y'})(\bar\str_{Z}+\bar\str_{Z'}) -\min\{\bar\str_X,\bar\str_{X'}\}\Big[\min\{\bar\str_Y,\bar\str_{Y'}\} \max\{\bar\str_Z,\bar\str_{Z'}\}\nn\\&+\max\{\bar\str_Y,\bar\str_{Y'}\} \min\{\bar\str_Z,\bar\str_{Z'}\}\Big]-\max\{\bar\str_X,\bar\str_{X'}\}\Big[\min\{\bar\str_Y,\bar\str_{Y'}\} \min\{\bar\str_Z,\bar\str_{Z'}\}\nn\\&+\max\{\bar\str_Y,\bar\str_{Y'}\} \max\{\bar\str_Z,\bar\str_{Z'}\}\Big] -2\min\{\bar\str_X,\bar\str_{X'}\}\min\{\bar\str_Y,\bar\str_{Y'}\} \min\{\bar\str_Z,\bar\str_{Z'}\} 
\nn\\=& (2-\str_X-\str_{X'})(2-\str_Y-\str_{Y'}) (2-\str_Z-\str_{Z'})  -r_{X}\Big( r_Y \ell_Z +\ell_Y r_Z\Big)\nn\\&  -  \ell_X\Big( r_Yr_Z +\ell_Y\ell_Z\Big)- 2r_Xr_Yr_Z,\nn \\
		=& K_{\max}.
		\label{pfthm2}
		\end{align}
		where $r_i=1-\max\{\str_i,\str_{i'}\}$, $\ell_i=1-\min\{\str_i,\str_{i'}\}$, and  $i=\{X,Y,Z\}$.
Here the third line is obtained by using the fact $\max\{a,c\}\max\{b,d\}+\min\{a,c\}\min\{b,d\} \geq ab+cd$ in second line  and the fourth line by verifying it for the case $\bar \str_X\leq\bar \str_{X'}, \bar \str_Y\leq\bar \str_{Y'}$ and  $\bar \str_Z\leq\bar \str_{Z'}$. 
\end{proof}
\section{Proof of Theorem \ref{fix-anglexnx} and Proposition \ref{fix-smax}}\label{opt_angle_merm}
The proof of Theorem \ref{fix-anglexnx} is as follows,
\begin{proof}
The Mermin operator for unbiased measurement observables is given by
\beq
\langle \mathcal{E} \rangle_{ub} =  \str_{Y}  \str_{Z} \Big[\str_X\bm x^{\top} T\left(\bm y\otimes\bm z'+  \bm y'\otimes \bm z\right) + \str_{X'} \bm x'^{\top}T\left(\bm y\otimes  \bm z-  \bm y'\otimes \bm z'\right)\Big],\nn 
\eeq
As it is always possible to find two orthogonal unit vectors $\bm{p}$ and $\bm{p}'$ such that
\begin{eqnarray}
\bm y\otimes\bm z'+  \bm y'\otimes \bm z=2 \cos \frac{\theta_{yz}}{2} \bm{p}, \qquad
\bm y\otimes  \bm z-  \bm y'\otimes \bm z'=2 \sin \frac{\theta_{yz}}{2} \bm{p}',
\end{eqnarray}
where $\theta_{yz}\in [0,\pi]$ is considered to be principal 
angle such that $\cos\theta_y\cos\theta_z=\cos\theta_{yz}$; $\theta_y$ and $\theta_z$ are the angles  
between $\bm{y}$ and $\bm{y}'$, and $\bm{z}$ and $\bm{z}'$ respectively.

The maximum value of Mermin operator is given by
\begin{eqnarray}
\EuScript{M} &&=2\str_{Y}\str_{Z} \max_{\substack{\bm{x},\bm{x}',\theta_{yz} ,\bm{p},\bm{p}'}} 
\Big |\str_X \bm{x}^{\top} T\bm{p}\cos \frac{\theta_{yz}}{2} +\str_{X'} \bm{x}'^{\top} T\bm{p}'\sin \frac{\theta_{yz}}{2}\Big |\nn\\&& \leq
2\str_{Y}\str_{Z}\max_{\theta_{yz},\bm{p},\bm{p}'} \Big[\str_{X}|T\bm{p}|\cos \frac{\theta_{yz}}{2} +\str_{X}'|T\bm{p}'|\sin \frac{\theta_{yz}}{2} \Big]\nn\\&&\leq
2\str_{Y}\str_{Z}\max_{\bm{p},\bm{p}'} \Big[\str^2_{X}|T\bm{p}|^2+\str^2_{X'}|T\bm{p}'|^2\Big]^{\frac{1}{2}}.
\end{eqnarray} 

The above inequality is achieved using the property $a\cos\theta+b\sin\theta\leq (a^2+b^2)^{\frac{1}{2}}$ and the second line is obtained by taking $\bm{x}=\frac{T \bm p}{|T \bm p|}$ and $\bm{x}'=\frac{T \bm p'}{|T \bm p'|}$, which yield the angle $\cos \theta_x=0$. 
Therefore, we can write
\beq
\EuScript{M} \leq 2\str_{Y}\str_{Z}\sqrt{\str_{X}^2s^2_1(T)+\str_{X'}^2 s^2_2(T)}.
\eeq
The bound is saturated by choosing  $\theta_{yz}$, s.t. it satisfies 
\beq\tan \frac{\theta_{yz}}{2}=\frac{\str_{X'}|T\bm{p}'|}{\str_X|T\bm{p}|}=\frac{ \str_{X'}s_2(T)}{\str_X s_1(T)}.
\eeq 
Finally, Eq. (\ref{tbound}) for arbitrary measurement on T-state can be analogously obtained, from Eq. (\ref{mop}), where $K_{\max}$ can be calculated using Eq. (\ref{kmax}) for $\str_X\geq \str_{X'}$, $\str_Y=\str_{Y'}$ and $\str_Z=\str_{Z'}$. Thus, we get
\beq
\EuScript{M} \leq 2\str_{Y}\str_{Z}\sqrt{\str_{X}^2s^2_1(T)+\str_{X'}^2 s^2_2(T)} +2(1-\str_{X})(1-\str_Y)(1-\str_Z).
\eeq
Hence, it is proved.
\end{proof}

The proof of Proposition \ref{fix-smax} is as follows,
\begin{proof}
Considering $s_1(T)=s_2(T)=s_{\max}(T)$ in Theorem \ref{thm3}, we have $\EuScript{M}_0=s_{\max}(T)I_+(V)$, where 
\begin{align}\label{iplussq}
	I^2_+(V)=\str_X^2(\str_Y^2\str_{Z'}^2+\str_{Y'}^2\str_Z^2)+\str_{X'}^2(\str_Y^2\str_Z^2+\str_{Y'}^2\str_{Z'}^2)+2\xi(\theta_x,\theta_y,\theta_z).
\end{align}
with the function given as
\begin{align}
\xi(\theta_x,\theta_y,\theta_z)=&a_1 \cos\theta_x \cos\theta_y+b_1\cos\theta_x \cos\theta_z +c_1 \cos\theta_y \cos\theta_z+ d_1\sin\theta_x[e_1\sin^2{\theta_y}\nn\\&+f_1\sin^2{\theta_z}+g_1 (1- \cos 2\theta_y \cos 2\theta_z)]^{\frac{1}{2}},
\end{align}
where $a_1=\str_X \str_{X'} \str_Y \str_{Y'} (\str_Z^2-\str_{Z'}^2)$, $b_1=\str_X \str_{X'} \str_Z \str_{Z'} (\str_Y^2-\str_{Y'}^2)$, $c_1=\str_Y \str_{Y'} \str_Z \str_{Z'} (\str_X^2-\str_{X'}^2)$, $d_1=\str_X \str_{X'}$, $e_1=\str_Y^2\str_{Y'}^2(\str_Z^4+\str_{Z'}^4)$, $f_1=(\str_Y^4+\str_{Y'}^4) \str_Z^2\str_{Z'}^2$ and $g_1=\str_Y^2 \str_{Y'}^2 \str_Z^2 \str_{Z'}^2$.
To maximize $I^2_+(V)$ is to maximize $\xi(\theta_x,\theta_y,\theta_z)$, which seems hard. 
Considering $\theta_z=\frac{\pi}{2}$, we end up with the following expression
\begin{align}\label{new-merm5}
\xi(\theta_x,\theta_y)=a_1 \cos\theta_x \cos\theta_y+ d_1\sin\theta_x[e_1\sin^2{\theta_y}+f_1+g_1 (1+\cos 2\theta_y)]^{\frac{1}{2}},
\end{align} 
Partial derivatives of $\xi$ with respect to $\theta_x$ and $\theta_y$ equal to zero, yields respectively,
\begin{align}
a_1\sin\theta_x\cos\theta_y=d_1\cos\theta_x[e_1\sin^2{\theta_y}+f_1+g_1 (1+\cos 2\theta_y)]^{\frac{1}{2}},\label{new-merm51}
\\
\sin{\theta_y}\Big(a_1 \cos\theta_x-\frac{d_1 [e_1-2g_1]\sin\theta_x\cos\theta_y}{[e_1\sin^2{\theta_y}+f_1+g_1 (1+\cos 2\theta_y)]^{\frac{1}{2}}}\Big)=0.\label{new-merm52}
\end{align} 
From Eq. (\ref{new-merm52}), we have,
\begin{align}\label{new-merm53}
\sin{\theta_y}=0 \quad {\rm or}\quad \cot \theta_x=\frac{d_1(e_1-2g_1)\cos\theta_y}{a_1[e_1\sin^2{\theta_y}+f_1+g_1 (1+\cos 2\theta_y)]^{\frac{1}{2}}}.
\end{align}
Using Eq. (\ref{new-merm53}) in Eq. (\ref{new-merm51}), we have first set of solutions,
\begin{align}\label{new-merm54}
\sin{\theta_y}=0 \quad {\rm and}\quad \tan \theta_x=\pm\frac{d_1 \sqrt{f_1+2g_1}}{a_1},
\end{align}
and second set of solutions exists if $a_1^2\neq d_1^2(e_1-2g_1)$, then 
\begin{align}\label{new-merm55}
\cos{\theta_y}=0 \quad {\rm and}\quad \cos \theta_x=0.
\end{align}
However, we find that $a_1^2= d_1^2(e_1-2g_1)$ is always true for our function $\xi(\theta_x,\theta_y)$, hence the second set of solutions does not hold for our case.
Maximizing $\xi$ using Eq. (\ref{new-merm54}), we get
\begin{align}\label{new-merm56}
\cos\theta_y = {\rm sign}(a_1),\qquad \tan \theta_x=\frac{d_1\sqrt{f_1+2g_1}}{|a_1|},\qquad \xi_1=\sqrt{a_1^2+d_1^2(f_1+2g_1)}.
\end{align}
Putting the values of $a_1,d_1, f_1$ and $g_1$, we get
\begin{align}\label{mopt_new}
\xi_1=\str_X \str_{X'}\sqrt{\str^2_Y \str^2_{Y'} (\str_Z^4+\str_{Z'}^4)+\str_Z^2 \str_{Z'}^2 (\str_Y^4+\str_{Y'}^4)}.
\end{align}
Note here that, because of the symmetry between $\theta_y$ and $\theta_z$ in $\xi$, we will have the same optimization (Eq. (\ref{mopt_new})) if we fix $\theta_y=\frac{\pi}{2}$ instead of $\theta_z$.

Furthermore, fixing $\theta_x=\frac{\pi}{2}$, we reach exactly the same optimization (Eq. (\ref{mopt_new})) of function $\xi$ as one of the solutions for the choices of $\theta_y=\theta_z=\frac{\pi}{2}$. The other three solutions in this case are always smaller than Eq. (\ref{mopt_new}). 

Hence, putting the value of $\xi_1$ in Eq. (\ref{iplussq}), we get the proposition  \ref{fix-smax}.
\end{proof}

\section{Proof of Theorem \ref{thm6} and Corollary \ref{corollary7}}\label{appen3}
The proof of Theorem \ref{thm6} goes as follows.
\begin{proof}
Similar to the proof of Theorem \ref{thm3}, we can write Eq. (\ref{svetbzero}) as
\begin{align}\label{svet-formap}
\langle \mathcal{E-E'} \rangle =\sum_{ijk}W_{i(jk)} \bm x_i^\top T (\bm y_j\otimes \bm z_k)={\rm Tr}[WM^\top],
\end{align}
where the matrix $W$ is defined as
\begin{align*}
		W=\begin{pmatrix}
		W_1 &\bm 0^{\top}&W_2& (0) \\
		\bm 0&0&\bm 0&\bm 0
		\end{pmatrix},\:\:\mbox{with}\:\:
W_1=\begin{pmatrix}
		A_+\cos\frac{\theta_x}{2}\cos\frac{\theta_y}{2}\cos\frac{\theta_z}{2}&  B_+\cos\frac{\theta_x}{2}\cos\frac{\theta_y}{2}\sin\frac{\theta_z}{2}  \\
		C_{+}\sin\frac{\theta_x}{2}\cos\frac{\theta_y}{2}\cos\frac{\theta_z}{2}& D_{+}\sin\frac{\theta_x}{2}\cos\frac{\theta_y}{2}\sin\frac{\theta_z}{2}
		\end{pmatrix},
\end{align*}
		\begin{align*}
{\rm and}\qquad		W_2=\begin{pmatrix}
		C_- \cos\frac{\theta_x}{2}\sin\frac{\theta_y}{2}\cos\frac{\theta_z}{2}
		& -D_- \cos\frac{\theta_x}{2}\sin\frac{\theta_y}{2}\sin\frac{\theta_z}{2} \\
		A_- \sin\frac{\theta_x}{2}\sin\frac{\theta_y}{2}\cos\frac{\theta_z}{2}&-B_-  \sin\frac{\theta_x}{2}\sin\frac{\theta_y}{2}\sin\frac{\theta_z}{2}
		\end{pmatrix},
		\end{align*}
		where $A_\pm, B_\pm, C_\pm,$ and $D_\pm$ are given by
		\begin{align}
	A_{\pm}&=  (\str_X \str_Y - \str_{X'}\str_{Y'})(\str_Z+\str_{Z'})\pm (\str_X \str_{Y'}+ \str_{X'}\str_Y )(\str_Z-\str_{Z'})  \nn \\
	B_{\pm}&=  (\str_X \str_{Y'}+ \str_{X'}\str_Y ) (\str_Z+\str_{Z'})\pm  (\str_X \str_Y - \str_{X'}\str_{Y'})(\str_Z-\str_{Z'})\nn \\
		C_{\pm}&=  (\str_X \str_Y + \str_{X'}\str_{Y'}) (\str_Z+\str_{Z'})\pm (\str_X \str_{Y'}- \str_{X'}\str_Y )(\str_Z-\str_{Z'})\nn \\
		D_{\pm}&=  (\str_X \str_{Y'}- \str_{X'}\str_Y )(\str_Z+\str_{Z'})\pm (\str_X \str_Y + \str_{X'}\str_{Y'})(\str_Z-\str_{Z'}), 
		\end{align}
and $M$ is the $3\times9$ matrix with coefficients
\beq
M_{i(jk)} := \bm x_i^\top T \bm y_j\otimes \bm z_k.
\eeq
The sub-matrices $W_1$ and $W_2$ contains the information about the strengths and relative angles of local measurements, whereas $M$ contains the global information of three body spin correlations. Using the Lemma \ref{lemma1} in Eq. (\ref{svet-formap}), and, as $s_3(W)=0$, we get the first equation of Theorem \ref{thm6}, i.e.,
\begin{align}\label{fnlbound2}
\EuScript{S}\leq   \sum_{i=1}^{2} s_i(T) s_i(W).
\end{align}
The optimal measurement directions can be determined using the relative angles and orthogonal rotation matrices $O_1,O_2$ and $O_3$.

To obtain the alternate expression in Theorem \ref{thm6}, we notice that $\EuScript{S}_0$ can be rewritten as 
\begin{align*}
\EuScript{S}_0=\half  [s_1(T)+s_2(T)]J_+(W) +  \half  [s_1(T)- s_2(T)]J_-(W),
\end{align*}
where $J_\pm(V)=s_1(W)\pm s_2(W)=\sqrt{w_+}\pm\sqrt{w_-}$, with $w_\pm$ as the eigenvalues of $W^\top W$. Using the identities $w_++w_-={\rm Tr}[W^\top W]$ and $w_+w_-={\rm det} (W^\top W)={\rm det} (W)^2$, one reaches 
\begin{align*}
J_\pm(W)^2={\rm Tr}[W^\top W]\pm 2|{\rm det} (W)|.
\end{align*}
Now explicit calculations of terms in the above equation will give us the expression of $J_\pm(W)$.
\end{proof}

In the following, we prove the Corollary \ref{corollary7}.
\begin{proof}	
	To obtain Eq. (\ref{corollary7eq}),  we first calculate the eigenvalues $\textsl{w}_\pm$ of $W^\top W$ for $\str_X=\str_{X'}$, $\str_Y=\str_{Y'}$ and $\str_Z=\str_{Z'}$,
	\begin{align} \label{wp}
	\textsl{w}_{+}= 8 \str_X^2 \str_Y^2 \str_Z^2 \left(1+ \cos \theta_y \cos \theta_z \right) \sin^2 \frac{\theta_x}{2},\nn\\
	\textsl{w}_{-}= 8 \str_X^2 \str_Y^2 \str_Z^2 \left(1- \cos \theta_y \cos \theta_z \right) \cos^2 \frac{\theta_x}{2}.
	\end{align}
	Applying Cauchy-Schwarz inequality to Eq. (\ref{fnlbound2}), we get
	\begin{align} \label{cor7prf}
	\EuScript{S} &\leq \left(\sum_{i=1}^{2} s^2_i(W)\right)^{1/2} \left(\sum_{i=1}^{2} s^2_i(T)\right)^{1/2}, \nn\\
	&= \sqrt{\textsl{w}_++\textsl{w}_-}\sqrt{s^2_1(T)+s^2_2(T)},\nn\\
	&=2\sqrt{2} \str_X\str_Y\str_Z \sqrt{1-\cos \theta_x \cos \theta_y \cos \theta_z}\sqrt{s^2_1(T)+s^2_2(T)},
	\end{align}
	where $\textsl{w}_+ =s^2_1(W), \textsl{w}_-=s^2_2(W)$ and equality holds if and only if  $\textsl{w}_+/\textsl{w}_-=s^2_1(T)/s^2_2(T)$.
	Using Eq. (\ref{wp}) we get the equality condition
	\beq \label{relan7p}
\cos \theta_y \cos \theta_z	 = \frac{s^2_1(T)-s^2_2(T)\tan^2 \frac{\theta_x}{2}}{s^2_1(T)+s^2_2(T)\tan^2 \frac{\theta_x}{2}}.
	\eeq
We notice that Eq. (\ref{cor7prf}) can be optimized even further. However, there exist many choices here. First, notice that $\theta_x\neq \pi$, as Eq. (\ref{relan7p}) becomes non physical. The choice $\theta_y=\pi$ and $\theta_z=0$ or, vice versa is also not possible as Eq. (\ref{relan7p}) is no longer satisfied. Therefore, the optimal solution is for $\theta_x=\frac{\pi}{2}$. Then, we reach to the solution, 
\begin{align}\label{new_find-q}
\EuScript{S} \leq
2\sqrt{2} \str_X\str_Y\str_Z \sqrt{s^2_1(T)+s^2_2(T)},
\end{align}
when the following conditions hold,
\beq 
\cos \theta_y \cos \theta_z	 = \frac{s^2_1(T)-s^2_2(T)}{s^2_1(T)+s^2_2(T)}, \quad {\rm and}\quad \theta_x=\frac{\pi}{2}.
\eeq
Notice also that the same bound (Eq. (\ref{new_find-q})) can be reached by considering $\cos\theta_y\cos\theta_z=0$, which will yield another equivalent condition from Eq. (\ref{relan7p}), 
\beq 
 \sin\theta_x=\frac{2s_1(T)s_2(T)}{s^2_1(T)+s^2_2(T)}, \quad {\rm and}\quad \cos\theta_y\cos\theta_z=0.
\eeq
Hence, we prove the result.
\end{proof}

\section{Proof of Theorem \ref{thm7}}\label{appen4}
\begin{proof}
	Following Eq. (\ref{constraint}), we see that $|\B|\leq\bar\str:= 1-\str$. We define, 
		$\B_X=\alpha\bar\str_X$, $\B_{X'}=\alpha'\bar\str_{X'}$, $\B_Y=\beta\bar\str_Y$, $\B_{Y'}=\beta'\bar\str_{Y'}$, $\B_Z=\gamma\bar\str_Z$, $\B_{Z'}=\gamma'\bar\str_{Z'}$, for suitable choices of $\alpha_,\alpha',\beta,\beta',\gamma,\gamma'=\pm1$ and $\nu=\gamma'/\gamma$.
		Substituting back in $L$, we get
		\begin{align}
		|L|\leq&\max_{\substack{\alpha,\alpha',\beta,\beta',\\\gamma,\gamma'=\pm1}} |\alpha\bar\str_X\left[\beta\bar\str_Y(\gamma\bar\str_Z+\gamma'\bar\str_{Z'}) + \beta'\bar\str_{Y'}(\gamma\bar\str_Z-\gamma'\bar\str_{Z'})\right] + \alpha'\bar\str_{X'}\Big[\beta\bar\str_Y(\gamma\bar\str_Z\nn \\&-\gamma'\bar\str_{Z'}) - \beta'\bar\str_{Y'}(\gamma\bar\str_Z+\gamma'\bar\str_{Z'})\Big]| \nn\\
		= &\max_{\nu=\pm1}\, (\bar\str_X\bar\str_Y-\bar\str_{X'}\bar\str_{Y'})|\bar\str_Z+\nu\bar\str_{Z'}| + (\bar\str_X\bar\str_{Y'}+\bar\str_{X'}\bar\str_Y)|\bar\str_Z-\nu\bar\str_{Z'}|\nn\\
		=& \max\{(\bar\str_X\bar\str_Y-\bar\str_{X'}\bar\str_{Y'}),(\bar\str_X\bar\str_{Y'}+\bar\str_{X'}\bar\str_Y)\}|\bar\str_Z+\bar\str_{Z'}| +\min\{(\bar\str_X\bar\str_Y\nn \\&-\bar\str_{X'}\bar\str_{Y'}),(\bar\str_X\bar\str_{Y'}+\bar\str_{X'}\bar\str_Y)\} |\bar\str_Z-\bar\str_{Z'}|\nn\\
		=&(\bar\str_X\bar\str_{Y'}+\bar\str_{X'}\bar\str_Y)|\bar\str_Z+\bar\str_{Z'}|+(\bar\str_X\bar\str_Y-\bar\str_{X'}\bar\str_{Y'})|\bar\str_Z-\bar\str_{Z'}|\nn\\
			=&\left[(1-\str_X)(1-\str_{Y'})+(1-\str_{X'})(1-\str_Y)\right]|2-\str_Z-\str_{Z'}|+\Big[(1-\str_X)\nn \\&\times(1-\str_Y)-(1-\str_{X'})(1-\str_{Y'})\Big]|\str_{Z'}-\str_Z|\nn\\
		=& L_{\max}.
		\label{pfthm9}
		\end{align}
	
Here the third line is obtained by using the fact $\max\{a,c\}\max\{b,d\}+\min\{a,c\}\min\{b,d\} \geq ab+cd$ in second line and the fourth line by verifying it for the case $\bar \str_X\leq\bar \str_{X'}, \bar \str_Y\leq\bar \str_{Y'}$. 
\end{proof}
\section{Proof of Theorem \ref{svet_maxeq} and Proposition \ref{fix-smaxsv}}\label{opt_angle_svet}
The proof of Theorem \ref{svet_maxeq} is as follows,
\begin{proof}
The unbiased Svetlichny operator for $\str_{Y}=\str_{Y'}$ and $\str_{Z}=\str_{Z'}$ is given by
	\begin{align} 
	\langle \mathcal{E}-\mathcal{E'} \rangle_{ub} =& \str_{Y}  \str_{Z} \Big[(\str_X \bm x^{\top}+\str_{X'} \bm x'^{\top})  T(\bm y\otimes  \bm z-  \bm y'\otimes \bm z')\nn\\&+ (\str_X \bm x^{\top}-\str_{X'} \bm x'^{\top}) T( \bm y\otimes\bm z'+  \bm y'\otimes \bm z)\Big].\nn
	\end{align}
	As it is always possible to find two orthogonal unit vectors $\bm{q}$ and $\bm{q}'$ such that
		\begin{eqnarray}
		\bm y\otimes  \bm z-  \bm y'\otimes \bm z' =2 \sin \frac{\theta_{yz}}{2} \bm{q}, \qquad
		\bm y\otimes\bm z'+  \bm y'\otimes \bm z= 2 \cos \frac{\theta_{yz}}{2}\bm{q}',
		\end{eqnarray}
where $\theta_{yz}\in [0,\pi]$ is considered to be principal 
angle such that $\cos\theta_y\cos\theta_z=\cos\theta_{yz}$; $\theta_y$ and $\theta_z$ are the angles between $\bm{y}$ and $\bm{y}'$, and $\bm{z}$ and $\bm{z}'$ respectively. 	
Let us consider new vectors, $\bm  q_\pm=( \sin \frac{\theta_{yz}}{2}\bm q  \pm \cos \frac{\theta_{yz}}{2}\bm q').$ 
Notice that the vectors $\bm  q_\pm$ becomes orthogonal unit vectors when $\theta_{yz}=\pi/2$. Hence, the maximum value of Svetlichny operator is given by
\begin{eqnarray}
\EuScript{S} &&=2\str_{Y}\str_{Z}\max_{\substack{\bm{x},\bm{x}',\bm{q_{\pm}}}} 
\Big |\str_X \bm{x}^{\top} T\bm{q_+} +\str_{X'} \bm{x}'^{\top} T\bm{q_-}\Big|\nn\\&&\leq
2\str_{Y}\str_{Z}\max_{\bm{q_{\pm}}} \Big[\str_{X}|T\bm{q_+}|+\str_{X'}|T\bm{q_-}|\Big]
\nn\\&&\leq 2 \str_{Y}\str_{Z}\Big[\str_{X}s_1(T)+\str_{X'}s_2(T)\Big],
\end{eqnarray} 
where we chose $\bm x= \frac{T \bm q_+}{|T \bm q_+|}$	and $\bm x'=\frac{T \bm q_-}{|T \bm q_-|}$, which yield the angle $\cos \theta_x=0$. 

Now considering the value of $L_{\max}$ for $\str_Y=\str_{Y'}$ and $\str_Z=\str_{Z'}$ in Eq. (\ref{lmax}), we have result for $T$-state.

Another way one can proceed considering the following equation with $\bm x_{\pm}=\str_X \bm{x}\pm\str_{X'} \bm{x}'$,
\begin{eqnarray}
	\EuScript{S} &&=2\str_{Y}\str_{Z}\max_{\substack{\bm{x},\bm{x}',\bm{q},\bm q'},\theta_{yz}} 
	\Big |\cos\frac{\theta_{yz}}{2}\bm x_-^\top T\bm{q} +\sin\frac{\theta_{yz}}{2}\bm x_+^{\top} T\bm{q'}\Big|
	\nn\\&&\leq
	2\str_{Y}\str_{Z}\max_{\bm{q},\bm q', \theta_{yz},\theta_x} \Big[|\bm x_-|\cos\frac{\theta_{yz}}{2}|T\bm{q}|+|\bm x_+|\sin\frac{\theta_{yz}}{2}|T\bm{q'}|\Big],\label{f4eq}
\end{eqnarray}
where $|\bm x_\pm|=\sqrt{\str^2_X+\str^2_{X'}\pm 2\str_{X}\str_{X'}\cos{\theta_x}}$. As $\theta_{yz}\in [0,\pi]$, all the terms in the above inequality are positive, then without loss of generality, considering $\cos\theta_x=0$.
\begin{eqnarray}
	\EuScript{S}&&= 2\str_{Y}\str_{Z}\sqrt{\str^2_X+\str^2_{X'}}\max_{\bm{q},\bm q', \theta_{yz}} \Big[\cos\frac{\theta_{yz}}{2}|T\bm{q}|+\sin\frac{\theta_{yz}}{2}|T\bm{q'}|\Big]
	\nn\\&&\leq
	2\str_{Y}\str_{Z}\sqrt{\str^2_X+\str^2_{X'}}\max_{\bm{q},\bm q'} \Big[|T\bm{q}|^2+|T\bm{q'}|^2\Big]^\frac{1}{2}
	\nn\\&&= 2 \str_{Y}\str_{Z}\sqrt{\str^2_X+\str^2_{X'}}\Big[s_1^2(T)+s_2^2(T)\Big]^\frac{1}{2},
\end{eqnarray} 
where the bound is achieved by choosing $\frac{\bm x_-}{|\bm x_-|}= \frac{T \bm q}{|T \bm q|}$	and $\frac{\bm x_+}{|\bm x_+|}=\frac{T \bm q'}{|T \bm q'|}$ and 
\beq 
\tan \frac{\theta_{yz}}{2}=\frac{ s_2(T)}{ s_1(T)},\:\:\mbox{and}\:\: \cos\theta_x=0.
\eeq
Similarly, one can choose $\cos\theta_x=1$ in Eq. (\ref{f4eq}), we get $|\bm x_\pm|=|\str_X\pm\str_{X'}|$ and $\bm x_{\pm}$ becomes parallel to $\bm x$ (equivalently to $\bm x'$). Therefore, in this case, we achieve optimality if the maximal singular value $s_{\max}(T)$ has degeneracy $2$,  then we have following bound, 
\begin{align}
	\EuScript{S}&= 2\str_{Y}\str_{Z}\max_{\bm{q},\bm q', \theta_{yz}} \Big[|\str_X-\str_{X'}|\cos\frac{\theta_{yz}}{2}|T\bm{q}|+|\str_X+\str_{X'}|\sin\frac{\theta_{yz}}{2}|T\bm{q'}|\Big]
\nn\\&\leq 2\str_{Y}\str_{Z}s_{\max}(T)\max_{\theta_{yz}} \Big[|\str_X-\str_{X'}|\cos\frac{\theta_{yz}}{2}+|\str_X+\str_{X'}|\sin\frac{\theta_{yz}}{2}\Big]	\nn\\&= 2\sqrt{2} \str_{Y}\str_{Z}s_{\max}(T)\sqrt{\str^2_X+\str^2_{X'}},
\end{align}
where the angles are
\beq 
\tan \frac{\theta_{yz}}{2}=\frac{ |\str_X+\str_{X'}|}{ |\str_X-\str_{X'}|},\:\:\mbox{and}\:\: \theta_x=0.
\eeq
Hence, the proof. Note however that the optimization above has not been done for arbitrary $\theta_x$.   
\end{proof}

The proof of Proposition \ref{fix-smaxsv} is as follows,
\begin{proof}
Considering $s_1(T)=s_2(T)=s_{\max}(T)$ in Theorem \ref{thm6}, we have $\EuScript{S}_0=s_{\max}(T)J_+(W)$, where 
\begin{align}\label{jplusapp}
J^2_+(W)=(\str_X^2+\str_{X'}^2)(\str_Y^2+\str_{Y'}^2)(\str_Z^2+\str_{Z'}^2)+2\zeta(\theta_x,\theta_y,\theta_z),
\end{align}
with the form of function being
\begin{align*}
\zeta(\theta_x,\theta_y,\theta_z)=&a_2\cos\theta_x+b_2\cos\theta_y+c_2\cos\theta_z-d_2 \cos\theta_x\cos\theta_y\cos\theta_z+e_2 \sin\theta_x\Big[f_2\sin^2\theta_y\\&+g_2\sin^2\theta_z+h_2(1- \cos 2\theta_y \cos 2\theta_z)\Big]^{\frac{1}{2}}
\end{align*}
where $a_2=\str_X \str_{X'}(\str_Y^2-\str_{Y'}^2)(\str_Z^2-\str_{Z'}^2)$, $b_2=\str_Y \str_{Y'} (\str_X^2-\str_{X'}^2)(\str_Z^2-\str_{Z'}^2)$, $c_2=\str_Z \str_{Z'}(\str_X^2-\str_{X'}^2)(\str_Y^2-\str_{Y'}^2)$, $d_2=4  \str_X \str_{X'}\str_Y \str_{Y'} \str_Z \str_{Z'}$, $e_2=2  \str_X \str_{X'}$, $f_2=\str_Y^2 \str_{Y'}^2 (\str_Z^4 +\str_{Z'}^4)$, $g_2=\str_Z^2 \str_{Z'}^2 (\str_Y^4+\str_{Y'}^4)$ and $h_2=\str_Y^2 \str_{Y'}^2\str_Z^2 \str_{Z'}^2$. 
 Using the same argument as above, we can choose $\sin\theta_x=0$, which will fix $\cos\theta_x={\rm sign}(a_2)$, then we end up with the following simplified function,
\begin{align}\label{solsv}
\chi(\theta_y,\theta_z):=\zeta(\theta_y,\theta_z)-|a_2|=b_2\cos\theta_y+c_2\cos\theta_z-d'_2 \cos\theta_y\cos\theta_z,
\end{align}
where $d'_2=d_2\cos{\theta_x}$.  Maximizing $\zeta$ function, will give us maximum $J_+(W)$.

Partial derivatives of $\chi$ with respect to $\theta_y$ and $\theta_z$ equal to zero, yields
\beq \label{dervsv}
\sin{\theta_y}(b_2-d'_2\cos\theta_z)=0 \quad {\rm and} \quad  \quad \sin{\theta_z}(c_2-d'_2\cos\theta_y)=0.
\eeq
which implies that the first set of solutions are
\beq \label{derv2sv}
\sin\theta_y=0 \quad {\rm and} \quad \sin\theta_z=0. 
\eeq
Maximizing $\chi$ under the constraints of Eq. (\ref{derv2sv}), we get
\beq \label{sol1sv}
\cos\theta_y = {\rm sign}(b_2),\qquad \cos\theta_z = {\rm sign}(c_2),\qquad \chi_1=|b_2|+|c_2|+|d'_2|,
\eeq 

On the other hand, Eq. (\ref{dervsv}) yield three sets of solutions which produces only one maxima, i.e.,
\begin{align} \label{sol2sv}
\sin\theta_y=0,\quad {\rm and}\quad \cos{\theta_z}=\frac{b_2}{d'_2};\quad {\rm or,}\quad
\cos{\theta_y}=\frac{c_2}{d'_2},\quad{\rm and}\quad\sin\theta_z=0;\nn\\
{\rm or,}\quad\cos{\theta_y}=\frac{c_2}{d'_2}, \quad {\rm and}\quad\cos{\theta_z}=\frac{b_2}{d'_2};\quad {\rm with} \quad \chi_2=\left|\frac{b_2c_2}{d_2'}\right|.\qquad
\end{align}
 Note that the solutions mentioned in above equation is valid only when it satisfies the following constraint
\beq \label{constraints_sv}
1\geq |\cos\theta_y\cos\theta_z|=\frac{|b_2c_2|}{d'^2_2},
\eeq
To check for global maxima, lets calculate the following identity,
\beq
\chi^2_1-\chi^2_2=\left(|b_2|+|c_2|+|d'_2|\right)^2-\frac{b^2_2c^2_2}{d'^2_2}\geq (|b_2|+|c_2|)^2+2(|b_2|+|c_2|)|d'_2|+d'^2_2\left(1-\frac{b^2_2c^2_2}{d'^4_2}\right)\geq 0,\nn
\eeq
as we find $\left(1-\frac{b^2_2c^2_2}{d'^4_2}\right)\geq 0$ from Eq. (\ref{constraints_sv}), which if satisfied, then whenever the first maxima exist, it is the global maxima. 

Substituting values of $b_2, c_2,$ and $d'_2$, in Eq (\ref{constraints_sv}) we get
\beq \label{consv}
1\geq \frac{(\str_X^2-\str_{X'}^2)^2| (\str_Y^2-\str_{Y'}^2)(\str_Z^2-\str_{Z'}^2)|}{16 \str^2_X \str^2_{X'} \str_Y \str_{Y'} \str_Z \str_{Z'}},
\eeq

 Eq. (\ref{consv}) can only be satisfied if we consider all the strengths to be nonzero. Therefore the best optimization can only be obtained using the first set of solutions, or in other words, $\chi_1$ helps to find the maximum violation of Svetlinchy inequality. This is also logical as it is never possible to violate Svetlichny inequality using (one of) the measurements with zero strength \cite{PhysRevLett.48.291}. 
Substituting values of $a_2, b_2, c_2$ and $d_2$, in Eq. (\ref{sol1sv}),  and using  Eq. (\ref{solsv}) we get 
\begin{align*}
\zeta_1=&\str_X \str_{X'}|\str_Y^2-\str_{Y'}^2||\str_Z^2-\str_{Z'}^2|+4  \str_X \str_{X'}\str_Y \str_{Y'} \str_Z \str_{Z'}\nn\\&+\Big(\str_Y \str_{Y'} |\str_Z^2-\str_{Z'}^2|+\str_Z \str_{Z'}|\str_Y^2-\str_{Y'}^2|\Big) |\str_X^2-\str_{X'}^2|.
\end{align*}
Hence, putting the above expression in Eq. (\ref{jplusapp}), we get the proposition  \ref{fix-smaxsv}.
\end{proof}

\section*{References}

\end{document}